\documentclass[preprint,12pt]{elsarticle}

\usepackage{geometry}
\usepackage{graphics,graphicx}
\usepackage[usenames,dvipsnames]{xcolor}
\usepackage{amsmath}
\usepackage{latexsym}
\usepackage{amsfonts}
\usepackage{mathrsfs}
\usepackage{bigints}
\usepackage{bm}
\usepackage{multirow}
\usepackage{amsthm}
\usepackage{amssymb}
\usepackage{caption}
\usepackage{subcaption}
\usepackage{float}
\usepackage[colorlinks,citecolor=blue,urlcolor=blue,bookmarks=false,hypertexnames=true]{hyperref}
\usepackage[title]{appendix}
\usepackage{tikz}
\usetikzlibrary{shapes.geometric, arrows}
\tikzstyle{startstop} = [rectangle, rounded corners, minimum width=3cm, minimum height=1cm,text centered, draw=black]
\tikzstyle{arrow} = [thick,->,>=stealth]
\graphicspath{ {Fig/} }

\usepackage[labelformat=simple]{subcaption}

\newtheorem{theorem}{Theorem}[section]

\usepackage{tikz}

\usetikzlibrary{shapes.geometric, arrows, shapes}
\tikzstyle{startstop} = [rectangle, rounded corners, minimum width=3cm, minimum height=1cm,text centered, draw=black]
\tikzstyle{arrow} = [thick,->,>=stealth]

\newcommand\solidrule[1][1cm]{\rule[0.5mm]{6mm}{2pt}}

\newcommand\dashdotrule{\mbox{%
\rule[0.5mm]{2.5mm}{2pt}\hspace{1mm}\rule[0.5mm]{1mm}{2pt}\hspace{1mm}\rule[0.5mm]{2.5mm}{2pt}}}

\definecolor{fadblu}{rgb}{0, 0.4470, 0.7410}
\definecolor{darkmar}{rgb}{0.6350, 0.0780, 0.1840}

\biboptions{sort&compress}

\makeatletter

\newcommand{\mathleft}{\@fleqntrue\@mathmargin0pt}
\newcommand{\mathcenter}{\@fleqnfalse}
\makeatother
\newlength{\bibitemsep}
\newlength{\bibparskip}
\let\oldthebibliography\thebibliography
\renewcommand\thebibliography[1]{%
  \oldthebibliography{#1}%
  \setlength{\parskip}{\bibitemsep}%
  \setlength{\itemsep}{\bibparskip}%
}

\usetikzlibrary{positioning, fit, calc}   
\tikzset{block/.style={draw, thick, text width=2cm ,minimum height=1.3cm, align=center},   
line/.style={-latex}     
}

\begin{document}

\begin{frontmatter}


\title{The role of inducible defence in ecological models: Effects of nonlocal intraspecific competitions}

\author[inst1]{Sangeeta Saha}
\ead{sangeetasaha629@gmail.com}
\author[inst1]{Swadesh Pal\corref{cor1}}
\ead{spal@wlu.ca}
\author[inst1,inst2]{Roderick Melnik}
\ead{rmelnik@wlu.ca}

\cortext[cor1]{Corresponding author}
\address[inst1]{MS2 Discovery Interdisciplinary Research Institute, Wilfrid Laurier University, Waterloo, Canada}
\address[inst2]{BCAM - Basque Center for Applied Mathematics, E-48009, Bilbao, Spain}

\begin{abstract}
Phenotypic plasticity is a key factor in driving the evolution of species in the predator-prey interaction. The natural environment is replete with phenotypic plasticity, which is the source of inducible defences against predators, including concealment, cave-dwelling, mimicry, evasion, and revenge. In this work, a predator-prey model is proposed where the prey species shows inducible defence against their predators. The dynamics produce a wide range of non-trivial and impactful results, including the stabilizing effect of the defence mechanism. The model is also analyzed in the presence of spatio-temporal diffusion in a bounded domain. It is found in the numerical simulation that the Turing domain shrinks with the increase of defence level. The work is extended further by introducing a nonlocal term in the intra-specific competition of the prey species. The Turing instability condition has been studied for the local model around the coexisting steady state, followed by the Turing and non-Turing patterns in the presence of the nonlocal interaction term. The work reveals how an increase in inducible defence reduces the Turing domain in the local interaction model but expands it when the range of nonlocal interactions is extended, suggesting a higher likelihood of species colonization.
\end{abstract}

\begin{keyword}
Phenotypic plasticity \sep Inducible defence \sep Predator-prey interactions \sep Nonlocal model \sep Hopf domains \sep Patterns in ecosystems
\end{keyword}

\end{frontmatter}


\section{Introduction} \label{sec:1}

Results from studies in the field of ecological and evolutionary dynamics have demonstrated the possibility of ecological and evolutionary processes coinciding \cite{hairston2005rapid, schoener2011newest}. The significance of phenotypic plasticity-based intraspecific variation in the ecology and evolution of species has come to light in recent times \cite{gangur2017changes}. In his work, American psychologist Baldwin affirmed how phenotypic plasticity influences adaptive evolution and aids in survival  \cite{baldwin1896new}. The term ``phenotypic plasticity" describes how an organism alters behaviour, morphology, and physiology when placed in a certain environment \cite{yamamichi2019modelling}. The predator-prey system, which serves as a model system for population dynamics and forms the foundation of the intricate food chain, food network, and biochemical network structure, has long been a key topic in ecology and biomathematics. Furthermore, theoretical studies concerning the predator-prey model are constantly being conducted \cite{pati2022delayed, liu2018dynamics, wu2016global, arsie2022predator}. The predator-prey system's phenotypic plasticity has a significant impact on the evolution of species. It is common in the natural world and includes induced defences against predation such as hiding, cave-dwelling, mimicry, evasion, counterattacking, etc. Inducible defences are an important ecological element that affects ecological dynamics, especially predator-prey stability, either directly or indirectly. \par

Currently, research on the function of inducible defences has mostly relied on studying and experimenting with particular biological systems. Inducible defences have a significant effect on predator-prey dynamics, as demonstrated by several empirical studies \cite{van2018inducible}. The striking effectiveness of a size-selective predator served as an initial framework for the inducible defence model that Riessen and Trevett-Smith developed \cite{riessen2009turning}. The model was subsequently verified using a model system of inducible defences: the induction of a neck spine in the water flea Daphnia in response to predatory larvae of the phantom midge Chaoborus. Other researchers also used experiments to study Daphnia's inducible defences \cite{boeing2010inducible, rabus2011growing}. Using experimental and field research, Zhang et al. \cite{zhang2017predator} investigated variations in the protective spine length of rotifers. To study how prey relies on chemical or visual clues to stay away from predators, several researchers conducted lab tests with an anuran tadpole and a predatory dragonfly nymph \cite{gomez2011invasive, takahara2012inducible}. Moreover, West et al. experimentally established that for a vulnerable animal, exposure to predators enhances anti-predator reflexes \cite{west2018predator}. \par

Theoretical approaches must be used to examine biological phenomena that experiments are unable to explain to fully investigate the mechanism of inducible defences. Yamamichi et al. proposed three modelling techniques to capture the whole spectrum of inducible defences in predator-prey models: Fitness Gradient (FG), Optimal Trait (OT), and Switching Function (SF) \cite{yamamichi2019modelling}. Also, to assess the impact of inducible defences on community persistence and stability, Vos et al. \cite{vos2004binducible} examined models of bitrophic and tritrophic food chains that include consumer-induced polymorphisms. A predator-prey model with anti-predator behaviour in the prey population was examined by Sun et al. \cite{sun2017predator}. According to this model, anti-predator behaviour occurs only when the prey population size exceeds a threshold. Using a prey (paramecium) and predator (planarians) experiment, Hammill et al. \cite{hammill2010predator} examined how inducible defences impact predator functional responses. A recent analysis of a prey-predator interaction in which the prey exhibits inducible protection against its predator due to infection was conducted by Liu and Liu \cite{liu2023dynamics}. Ramos-Jiliberto et al. \cite{ramos2007pre} considered the general Holling type II functional response function as
$$\phi_{i}=\frac{1}{H_{i}+(A_{i}N_{i-1})^{-1}}$$
with $A_{i}$ and $H_{i}$ as the attack rate and the handling time of a prey item, respectively, which, after implementation of the inducible defence factor, takes the following form (parameters are given in the article): 
$$\phi_{y}=\frac{1}{h_{y}[1+(E_{y}-1)D]+\{a_{y}[1+(F_{y}-1)D]x\}^{-1}}.$$
Through numerical simulations, they examined the system's qualitative characteristics. \par

A conceptual model was proposed by Gonz$\Acute{a}$lez-Olivares et al. \cite{gonzalez2017multiple} to categorize various types of prey defences. There, it is believed that under pressure from predators, the defensive characteristics of the prey encompass modifications to their morphology, behaviour, physiology, or life history. On the contrary, as top predators have non-consumptive impacts, the predator's defensive traits will lower the predation rate \cite{gonzalez2017multiple}. The functional response, assuming $U$ and $V$ as the corresponding prey and predator biomass, has the form $p(U)=qU(1-R)$. The defensive trait set is denoted by the parameter $R= U_{r}/U$, where $U_{r}$ is the biomass of prey that is resistant to predators. Depending on the sensitivity of $R$ to prey and predator biomass, the authors developed a six-type categorization of prey defences. Along with it, they have taken into consideration that $\partial R/\partial U=0$ and $\partial R/\partial V>0$ hold, which indicates that the average prey's resilience to predators is a function of the biomass of predators. We refer to this type of protective reaction as inducible defences. \par

Now, the homogeneous distribution throughout the whole spatial domain is an assumption made by temporal models of interacting populations. However, it is hard to correctly model the random movement of individuals across the short or long-range without taking spatio-temporal models into account. The emergence of spatio-temporal patterns in predator-prey interactions stemming from the reaction-diffusion equation has garnered significant interest since Alan Turing's pioneering research on chemical morphogenesis \cite{turing1990chemical}. Through the diffusion terms within their habitats, spatio-temporal models can capture the impact of interacting species moving in arbitrary directions. When species are heterogeneously dispersed throughout their habitats in ecology and form local patches, reaction-diffusion systems describe a wide range of dynamic properties that the species show \cite{ZHANG20111883, pal2023role}. The kinetics of the reaction play an important role in developing different kinds of stationary and non-stationary spatial patterns. The reaction-diffusion systems may also be used to explain models of ecological invasions \cite{zhang2024spreading}, the spread of diseases \cite{wilson1997analysis}, etc. Researchers have extensively examined the spatio-temporal pattern formation in reaction-diffusion models and have identified several mechanisms, including Hopf bifurcation \cite{fussmann2000crossing}, diffusion-driven instabilities (Turing instability) \cite{PhysRevE.75.051913}, etc. Non-stationary spatial patterns among these comprise spatio-temporal chaotic patterns, periodic travelling waves, and travelling waves. The population patches never settle down to any stationary distribution due to these later patterns, which continue to change over time. Most of the stationary patterns are seen when the parameter values are chosen from the Turing or Turing-Hopf domain. Non-stationary spatial patterns, such as travelling waves, modulated travelling waves, etc., are also seen for a specific range of parameter values, which are not connected to Turing instability. \par

The spatio-temporal local models are formulated based on the assumption that the individuals present at a given spatial point interact with one other only and that members of both species can move between spatial points. However, this kind of equation with local interaction can not model the situation in which an individual placed at one spatial position can access resources located at another spatial point. There are some published works where nonlocal interaction factors are incorporated into the reaction kinetics to describe such scenarios \cite{BANERJEE20172, BAYLISS201737, pal2018analysis}. In a nonlocal predator-prey interaction, individuals at a spatial position can migrate to a different place to use the resources located at that location. Researchers have shown that the nonlocal model produces a spatio-temporal chaotic pattern for a large consumption rate when the predator starts to move to a nearby location in the presence of abundant prey \cite{BANERJEE20172}. Also, the same model with local interaction does not produce any Turing pattern, but the nonlocal model does. Now, spatio-temporal models are usually considered under various types of boundary conditions. No-flux and periodic boundary conditions are well-known and can be justified in ecology. Different types of stationary and non-stationary patterns are observed with periodic and Neumann boundary conditions. The Turing pattern is an example of a stationary pattern, and non-stationary patterns include travelling waves, wavetrains, spatio-temporal chaos, etc \cite{guan2011spatiotemporal, baurmann2007instabilities}. The stationary patterns are important to determine the long-term dynamics of the species in a habitat. The introduction of the nonlocal interaction gives rich dynamics compared to the local model. For the nonlocal model, Turing patterns are found in one-dimension with periodic boundary conditions \cite{pal2018analysis, segal2013pattern, ninomiya2017reaction}. In this case, the kernel function is assumed to be periodic with finite support. Many types of kernel functions with finite or infinite support are available in literature \cite{pal2018analysis, segal2013pattern, ninomiya2017reaction, 
sherratt2016invasion, genieys2006pattern, fuentes2003nonlocal, bayliss2017complex}. Here, we choose the top-hat kernel function, and the kernel functions are chosen so that the homogeneous steady-states for the nonlocal model remain the same as that of the local model. Let $\sigma$ be the range of nonlocal interaction which represents the maximum distance on each side of its spatial location in which a species can access resources. Due to periodic boundary conditions, a species always accesses resources in a region of length $2\sigma$. \par


Ecological studies focus only on relationships between organisms and their natural surroundings. For population ecology, their interactions are crucial, and it is via these interactions that ecological systems develop. In the natural world, resources are often dispersed widely throughout the ecosystem. Therefore, due to survival and the need to find food, organisms move about (diffusiveness) within the ecosystem. Species mobility has a significant impact on their trophic interactions, which have led to the evolution of various spatial patterns in nature. There's ample evidence that terrestrial and aquatic populations may establish patterns \cite{chang2016spatial, wang2004stationary}. Numerous factors, such as stochastic processes, environmental changes, deterministic processes, species expansion, mobility, and so on, all play an important part in forming spatial patterns. Patterns are prevalent in the ecological system owing to species interaction among habitats \cite{melese2011pattern}. Interacting organisms have a deterministic process that leads to self-organized spatial patterns. \par

The distributions of populations, being heterogeneous, depend not only on time but also on the spatial positions in the habitat. So, it is natural and more precise to study the corresponding PDE problem.  
The spatio-temporal predator-prey dynamics is given as follows:
\mathcenter
\begin{equation}\label{eq:int1}
\begin{aligned} 
\frac{\partial U(\Tilde{\textbf{x}},T)}{\partial T} &= D_{U}\Delta U+RU\left(1-\frac{U}{K}\right)-f(U,V)V,\ \ \ \mbox{in}\  \Omega_{T}:=\Omega\times(0,T), \\
\frac{\partial V(\Tilde{\textbf{x}},T)}{\partial T} &= D_{V}\Delta V+\xi_{1}f(U,V)V-\delta_{1}V,\ \ \ \mbox{in}\ \Omega_{T}, \\
&U(\Tilde{\textbf{x}},0)=U_{0}, \ \ V(\Tilde{\textbf{x}},0)=V_{0}, \ \   \Tilde{\textbf{x}}\in \Omega,  \\
\end{aligned}
\end{equation}
where $U(\Tilde{\textbf{x}},T)$ and $V(\Tilde{\textbf{x}},T)$ represent the densities of the prey species and predator species
at spatial location $\Tilde{\textbf{x}}$ and time $T$, respectively, and $f(U, V)$ are chosen as different kinds of functional responses depending on the corresponding analysis. The parameters $R$ and $K$ represent the prey's intrinsic growth rate and carrying capacity in the habitat. The predator species' biomass conversion rate is $\xi_{1}$, whereas their natural death rate is represented by the parameter $\delta_{1}$. The parameters $D_{U}$ and $D_{V}$ indicate the self-diffusion of the prey and predator species, respectively. The diffusion strategy is defined as the random relocation of organisms from higher concentration to lower concentration areas. We assume the domain $\Omega$ is a bounded and open subset of $\mathbb{R}^{d}$, $d \leq 3$, with a smooth boundary $\partial\Omega$  and $\Delta$ denotes $\displaystyle\Sigma_{i=1}^{d}\partial^{2}/\partial x_{i}^{2}$. \par

This article concentrates on a predator-prey system where the functional response is of Holling type II. In addition, we consider an inducible defence strategy for the prey species toward their predator. In this case, the Holling type II functional response is modified into 
$$f(U, V)=\frac{MU(1-\Psi(V))V}{1+Mh(1-\Psi(V))U}$$ 
where $\Psi(V)=\omega V/(\Phi+V)$ with $M$ is the predator's consumption rate, $h$ is the handling time of the predator, $\omega$ is the rate of defence by the prey population, and $\Phi$ is a positive constant. After implementing the inducible defence into the reaction-diffusion system (\ref{eq:int1}), it becomes
\mathcenter
\begin{equation}\label{eq:det0}
\begin{aligned} 
\frac{\partial U(\Tilde{\textbf{x}},T)}{\partial T} &= D_{U}\Delta U+RU\left(1-\frac{U}{K}\right)-\frac{MU(1-\Psi(V))V}{1+Mh(1-\Psi(V))U}, \\
\frac{\partial V(\Tilde{\textbf{x}},T)}{\partial T} &= D_{V}\Delta V+\frac{\xi_{1}MU(1-\Psi(V))V}{1+Mh(1-\Psi(V))U}-\delta_{1}V,
\end{aligned}
\end{equation}
defined in a bounded domain $\Omega\subset \mathbb{R}^{n}$ with smooth boundary, supplemented with non-negative initial and periodic boundary conditions.

Dealing with a non-dimensional version of the model \eqref{eq:det0} can simplify the study by reducing the number of parameters, and thus, considering the non-dimensional species densities, space, and time variables as 
$$U=Ku, V=(hKR)v, \Tilde{\textbf{x}}=\frac{1}{\sqrt{R}}\textbf{x},~\mbox{and}~ T=\frac{t}{R}.$$
Substituting these non-dimensional variables into the system \eqref{eq:det0} and dropping the tildes on the dimensionless variable $\Tilde{\textbf{x}}$, we obtain:
\mathcenter
\begin{equation}\label{eq:diff1}
\left\{ \begin{aligned} 
\frac{\partial u(\textbf{x},t)}{\partial t} &= d_{1}\Delta u+u(1-u)-\frac{\left(1-\psi(v)\right)uv}{m+\left(1-\psi(v)\right)u}, \\
\frac{\partial v(\textbf{x},t)}{\partial t} &= d_{2}\Delta v+\frac{\xi\left(1-\psi(v)\right)uv}{m+\left(1-\psi(v)\right)u}-\delta v, \\
\end{aligned}\right.
\end{equation}
where $\psi(v)=\Psi(V/(hKR))\equiv\omega v/(\phi+v)$ with the same initial and boundary conditions as mentioned earlier. The system parameters are chosen to be positive in the system where $m=1/MhK,\ \phi=\Phi/hKR,\ \xi=\xi_{1}/hR$ and $\delta=\delta_{1}/R$. We have considered the parametric restriction $0\leq \omega<1$ to maintain biological relevance.

Here, a predator-prey relationship is presented in which the prey applies inducible defence strategy towards their predator species. The main focus here is to analyze the effect of defence on the population. However, we have not restricted ourselves to the analysis of the local model, but we have also explored how the incorporation of the nonlocal fear term affects the overall dynamic behaviour. In the following sections of the manuscript, we have organized as follows: In Section \ref{sec:2}, the temporal model is proposed, and its local dynamical behaviour is analyzed. In Section \ref{sec:3}, the one-dimensional diffusion terms are incorporated into the system under periodic boundary conditions, and the local stability of the spatio-temporal model is studied. The model is enhanced with nonlocal interactions in Section \ref{sec:4}, and stability analysis for the corresponding system has been performed. All the analytical findings have been validated through numerical simulations in Section \ref{sec:5}, and the work ends in Section \ref{sec:6} with an overall conclusion. 

\section{Dynamics of the temporal model} \label{sec:2}
First, we analyze the dynamics of predator-prey interaction in the absence of species diffusion, which takes the form
\begin{equation}\label{eq:det1}
\begin{aligned}
\frac{du}{dt}&=u\left[(1-u)-\frac{\left(1-\frac{\omega v}{\phi+v}\right)v}{m+\left(1-\frac{\omega v}{\phi+v}\right)u}\right]=u\psi_{1}(u,v),\ \ u(0)\geq 0 \\
\frac{dv}{dt}&=v\left[\frac{\xi\left(1-\frac{\omega v}{\phi+v}\right)u}{m+\left(1-\frac{\omega v}{\phi+v}\right)u}-\delta \right]=v\psi_{2}(u,v), \ \ v(0) \geq0.
\end{aligned}  
\end{equation} 
The system parameters are positive and are the same as those described for model \eqref{eq:diff1}. Moreover, the parametric restriction $0\leq \omega<1$ holds. 

\subsection{Positivity and boundedness of system (\ref{eq:det1})}\label{subsec-2.1}

Let us denote, $\Upsilon=\{(u,v)\in \mathbb{R}^{2}_{+}| u\geq 0, v\geq 0\}$ on the basis of ecological significance. We intend to show that the proposed model (\ref{eq:det1}) is well-posed as the system variables are positive and bounded. The right-hand side of the system (\ref{eq:det1}) is continuously differentiable and locally Lipschitzian (as they are polynomials and rationals in $(u,v)$) in the first quadrant $\Upsilon$. Therefore, a unique solution $(u(t),v(t))$ of the system with positive initial conditions $(u(0), v(0))$  exist on the interval $[0,\infty)$ according to Theorem A.4 in \cite{thieme2018mathematics}. From the first and second equation of (\ref{eq:det1}) we have
\begin{align*}
u(t)&=u(0)\exp\left[\int^t_0 \psi_{1}(u(z),v(z))\,dz\right]> 0, \ \textrm{for} \ u(0) \geq 0, \\
\textrm{and}\ \ v(t)&=v(0)\exp\left[\int^t_0 \psi_{2}(u(z),v(z))\,dz\right]> 0, \ \textrm{for} \ v(0)\geq 0.
\end{align*}
So, the solutions of the system (\ref{eq:det1}) remain feasible with time. Now, we prove, the solutions, starting in $\mathbb{R}_{+}^{2}$ are uniformly bounded too for $t>0$.
The first equation of (\ref{eq:det1}) gives:
\begin{equation*}
\begin{aligned}
\frac{du}{dt}&=u(1-u)-\frac{\left(1-\frac{\omega v}{\phi+v}\right)uv}{m+\left(1-\frac{\omega v}{\phi+v}\right)u}\leq u(1-u) \\
\displaystyle \Rightarrow & \limsup_{t\rightarrow \infty}u(t)\leq 1.
\end{aligned}
\end{equation*}

\noindent Consider, $W(t)= u(t)+\frac{1}{\xi}v(t)$. Then we have
\begin{equation*}
\begin{aligned}
\frac{dW}{dt}= \left(\frac{du}{dt}+\frac{1}{\xi}\frac{dv}{dt}\right)&=[u(1-u)+u]-u-\frac{\delta v}{\xi} \leq 1-\tau W, \ \ \ \mbox{where}\ \tau=\min\{1,\delta\}.
\end{aligned}
 \end{equation*}
Now, using the comparison theorem on the above differential inequality, we have
$$\displaystyle 0\leq W(t)\leq \frac{1}{\tau}+ \left(W(0)-\frac{1}{\tau}\right)\exp(-\tau t)$$ where $W(0)=W(u(0),v(0))$. As $\displaystyle t\rightarrow\infty,\ 0<W(t)\leq 1/\tau+\epsilon$ for sufficiently small $\epsilon>0$. Henceforth, the solutions of system (\ref{eq:det1}), initiating from the positive initial condition, enter into the region: 
$$\mathcal{T} = \{(u,v)\in \mathbb{R}^{2}_{+}: 0<u(t)\leq1; 0\leq W(t)\leq 1/\tau+\epsilon, \epsilon>0\}.$$

\subsection{Equilibrium points and local stability analysis of system (\ref{eq:det1})}\label{subsec-2.2}
Solving the prey and predator nullclines, we obtain that the temporal system has a trivial equilibrium point $E_{0}=(0,0)$, a predator-free equilibrium point $E_{1}=(1,0)$, and an interior equilibrium point $E^{*}=(u^{*},v^{*})$ where $\displaystyle u^{*}=\xi m(\phi+v^{*})/[(\xi-\delta)\{\phi+(1-\omega)v^{*}\}]$ and $v^{*}$ is the root of the following equation 
\begin{align}\label{eq:2.3}
   A_{1}v^{3}+A_{2}v^{2}+A_{3}v+A_{4}=0,
\end{align}
where $A_{1} =\delta(\xi-\delta)^{2}(1-\omega)^{2}, A_{2} =2\phi\delta(1-\omega)(\xi-\delta)^{2}+\xi\omega\delta^{2}m^{2}-\xi\delta m(1-\omega)(\xi-\delta-m\delta), A_{3} =\delta\phi^{2}(\xi-\delta)^{2}+\xi\phi\omega\delta^{2}m^{2}-m\xi\delta\phi(2-\omega)(\xi-\delta-m\delta),$ and $A_{4} =-m\xi\delta\phi^{2}(\xi-\delta-m\delta)$.

Now, the feasibility of $u^{*}$ holds when $\xi>\delta$ and $v^{*}>0$. On the other hand, if $\xi\leq (1+m)\delta$, then (\ref{eq:2.3}) does not have any positive roots as all of $A_{i}$ becomes positive in this case. It indicates that the species cannot coexist in the environment under this parametric restriction. But if $\xi>(1+m)\delta$, then by Descartes's rule, it has at least one or at most three positive roots depending on $A_{2}$ and $A_{3}$. 

The local stability criterion of the equilibrium points can be determined by analyzing the eigenvalues of corresponding Jacobian matrices. The Jacobian matrix of system (\ref{eq:det1}) at an equilibrium $E(u,v)$ is 
\mathcenter
\begin{equation}\ \label{eq:det2}
\textbf{J}=  \begin{pmatrix}
a_{11} & a_{12} \\
a_{21} & a_{22} 
\end{pmatrix},
\end{equation}
where $\displaystyle a_{11}=1-2u-\frac{m(1-\psi(v))v}{[m+(1-\psi(v))u]^{2}},\ a_{12}=-\frac{u(1-\psi(v))\{m+(1-\psi(v))u\}-muv\psi'(v)}{[m+(1-\psi(v))u]^{2}},  \\
a_{21}=\frac{\xi m(1-\psi(v))v}{[m+(1-\psi(v))u]^{2}},\ a_{22}=\frac{\xi u[(1-\psi(v))\{m+(1-\psi(v))u\}-mv\psi'(v)]}{[m+(1-\psi(v))u]^{2}}-\delta$. 

The Jacobian matrix at $E_{0}$ gives the eigenvalues as $1$ and $-\delta$ which makes the equilibrium an unstable point. As for the predator-free equilibrium $E_{1}$, the Jacobian matrix admits two eigenvalues as $\left(-1, \xi/(m+1)-\delta\right)$. So, $E_{1}$ is locally asymptotically stable (LAS) when $\xi<\delta(1+m)$ holds. Otherwise, it becomes a saddle point. The Jacobian matrix at $E^{*}$ is given by

\begin{equation*}
\textbf{J}(E^*)=\textbf{J}|_{E^{*}}=\begin{pmatrix}
  a_{11} & a_{12} \\
a_{21} & a_{22}  
\end{pmatrix},
\end{equation*}
where $\displaystyle a_{11}=-u^{*}+\frac{\{\phi+(1-\omega)v\}^{2}uv}{[m(\phi+v)]^{2}}, \\ a_{12}=\frac{-m(1-\omega)uv(\phi+v)-\{\phi+(1-\omega)v\}u[(1-\omega)uv+\phi(m+u)]}{[m(\phi+v)+\{\phi+(1-\omega)v\}u]^{2}} , \\ a_{21}=\frac{\xi mv(\phi+v)[\phi+(1-\omega)v]}{[m(\phi+v)+\{\phi+(1-\omega)v\}u]^{2}}$ and $\displaystyle a_{22}=\frac{-m\omega\xi\phi uv}{[m(\phi+v)+\{\phi+(1-\omega)v\}u]^{2}}$. \\
The characteristic equation corresponding to $\textbf{J}|_{E^{*}}$ is given as follows: 
\mathcenter
\begin{equation}\label{eq:2.5}
\lambda^{2}+C_{1}\lambda+C_{2}=0,
\end{equation}
where $C_{1}=-\mbox{tr}(\textbf{J}(E^{*}))=-(a_{11}+a_{22})$ and $C_{2}=\det(\textbf{J}(E^{*}))=a_{11}a_{22}-a_{12}a_{21}$. For satisfying the local stability condition of Routh-Hurwitz criteria, $C_{1}>0$ and $C_{2}>0$ need to be satisfied, which implies the conditions
\begin{equation} \label{eq:2.6}
    a_{11}+a_{22}<0\ \textrm{and}\ a_{11}a_{22}>a_{12}a_{21}
\end{equation}
have to be held.

\subsection{Temporal bifurcations around the equilibrium points}\label{subsec-2.3}

The temporal bifurcations around the equilibrium points are analyzed mainly with the help of Sotomayor's theorem and Hopf's bifurcation theorem \cite{perko2013differential}. In the system, if the stability condition of any of the equilibrium points violates in such a way that the corresponding determinant becomes $0$, giving a simple zero eigenvalue, then there will occur transcritical bifurcation, and we can observe the exchange of stability in that bifurcation threshold. To verify the conditions of Sotomayor's theorem, the Jacobian matrix at the bifurcating equilibrium point has to contain a simple zero eigenvalue. Let $\textbf{V}= (v_{1},v_{2})^{T}$ and $\textbf{W}= (w_{1},w_{2})^{T}$, respectively, be the eigenvectors of the corresponding Jacobian matrix and its transpose for a zero eigenvalue at the equilibrium point. Let, $\Psi=(\psi_{1},\psi_{2})^{T},$ where 
\begin{align*}
 \psi_{1}=u(1-u)-\frac{\left(1-\frac{\omega v}{\phi+v}\right)uv}{m+\left(1-\frac{\omega v}{\phi+v}\right)u}, \ \ \
 \psi_{2}=\frac{\xi\left(1-\frac{\omega v}{\phi+v}\right)uv}{m+\left(1-\frac{\omega v}{\phi+v}\right)u}-\delta v.
\end{align*}

\begin{theorem}
System (\ref{eq:det1}) undergoes a transcritical bifurcation around $E_{1}(1,0)$ at $\displaystyle m_{tc}=\xi/\delta-1$, choosing $m$ as the bifurcating parameter.
\end{theorem}
\begin{proof}
The Jacobian matrix corresponding to $E_{1}$ is given by
\mathcenter
\begin{equation*}
\textbf{J}|_{E_{1}}=\begin{pmatrix}
-1 & \frac{1}{m+1} \\
0 & \frac{\xi}{m+1}-\delta
\end{pmatrix},
\end{equation*}

The eigenvalues are given by $\left(-1, \xi/(m+1)-\delta \right)$, which are negative because of the stability conditions of the equilibrium point. Let $m_{tc}$ be the value of $m$ such that $\xi=\delta(m+1)$ so that $\textbf{J}|_{E_{1}}$ has a simple zero eigenvalue at $m_{tc}$. Now, the calculations give $\textbf{V}=(1, m+1)^{T}$ and $\textbf{W}=(0,1)^{T}.$ Therefore, we have
\begin{equation*}
\begin{aligned}
\Omega_{1}&= \textbf{W}^{T}\cdot\Psi_{m}(E_{1}, m_{tc})=\frac{-\xi uv(\phi+v)[\phi+(1-\omega)v]}{[m(\phi+v)+\{\phi+(1-\omega)v\}u]^{2}}\bigg|_{E_{1}}=0, \\
\Omega_{2}&= \textbf{W}^{T}\left[D\Psi_{m}(E_{1}, m_{tc})\textbf{V}\right]=-\frac{\xi}{m+1}\neq0, \\
\textrm{and}\ \Omega_{3}&= \textbf{W}^{T}\left[D^{2}\Psi(E_{1},m_{tc})(\textbf{V},\textbf{V})\right]=\frac{2m\xi\{\phi-\omega(m+1)\}}{\phi(m+1)}=\frac{2(\xi-\delta)(\phi\delta-\xi\omega)}{\phi\delta} \neq0. \\
\end{aligned}
\end{equation*}
\mathcenter
Hence, by Sotomayor's Theorem, the system undergoes a transcritical bifurcation around $E_{1}$ at $m=m_{tc}$.
\end{proof}

If any of the mentioned inequalities in (\ref{eq:2.6}) is violated, the equilibrium point becomes unstable, and the system performs oscillatory or non-oscillatory behaviour. The system starts to oscillate around $(u^{*}, v^{*})$ if $C_{1}>0$ along with $C_{1}^{2}-4C_{2}<0$ as the eigenvalues will be in the form of the complex conjugate in this case. So, we get the following theorem:

\begin{theorem}
If $E^{*}$ exists with the feasibility conditions, then a simple Hopf bifurcation occurs at unique $m=m_{H}$, where $m_{H}$ is the positive root of $C_{1}(m)=0$, providing $C_{2}(m_{H})>0$ (stated in equation (\ref{eq:2.5})).
\end{theorem}
\begin{proof}
At $m=m_{H}$, the characteristic equation of system (\ref{eq:det1}) at $E^{*}$ is $(\lambda^{2}+C_{2})=0$, and so the equation has a pair of purely imaginary roots $\lambda_{1}=i\sqrt{C_{2}}$ and $\lambda_{2}=-i\sqrt{C_{2}}$ where $C_{2}(m)$ is a continuous function of $m$. \\
In the small neighbourhood of $m_{H},$ the roots are $\lambda_{1}=p_{1}(m)+ip_{2}(m)$ and $\lambda_{2}=p_{1}(m)-ip_{2}(m)\ (p_{1},\ p_{2}\in \mathbb{R}$). \\
To show the transversality condition, we check $\displaystyle \left(\frac{d}{dm}[Re(\lambda_{i}(\omega))]\right)\bigg|_{m=m_{H}}\neq 0,$ for $i=1,2.$ \\
Put $\lambda(m)=p_{1}(m)+ip_{2}(m)$ in (\ref{eq:2.5}), we get
\mathcenter
\begin{equation} \label{eq:2.7}
(p_{1}+ip_{2})^{2}+C_{1}(p_{1}+ip_{2})+C_{2}=0.
\end{equation}
Differentiating with respect to $m$, we get
\begin{equation*}
2(p_{1}+ip_{2})(\dot{p_{1}}+i\dot{{p_{2}}})+C_{1}(\dot{p_{1}}+i\dot{p_{2}})+\dot{C_{1}}(p_{1}+ip_{2})+\dot{C_{2}}=0.
\end{equation*}
Comparing the real and imaginary parts from both sides, we have 
\begin{subequations}\label{eq:2.8}
\begin{align} 
(2p_{1}+C_{1})\dot{p_{1}}-(2p_{2})\dot{p_{2}}+(\dot{C_{1}}p_{1}+\dot{C_{2}}) = 0, \label{eq:2.8a} \\
(2p_{2})\dot{p_{1}}+(2p_{1}+C_{1})\dot{p_{2}}+\dot{C_{1}}p_{2}=0. \label{eq:2.8b}
\end{align}
\end{subequations}
Solving we get, $\displaystyle \dot{p}_{1}=\frac{-2p_{2}^{2}\dot{C_{1}}-(2p_{1}+C_{1})(\dot{C_{1}}p_{1}+\dot{C_{2}})}{(2p_{1}+C_{1})^{2}+4p_{2}^{2}}$. \\
At, $p_{1}=0,\ p_{2}=\pm \sqrt{C_{2}}:\ \displaystyle \dot{p}_{1}=\frac{-2\dot{C_{1}}C_{2}-C_{1}\dot{C_{2}}}{C_{1}^{2}+4C_{2}}\neq 0$. Hence, this completes the proof.
\end{proof}




\section{Linear stability analysis of the spatio-temporal model around $(u^{*},v^{*})$}\label{sec:3}

We intend to find the condition for Turing instability. If the homogeneous steady state of the temporal model is locally stable to infinitesimal perturbation but becomes unstable in the presence of diffusion, a scenario of Turing instability occurs. A necessary condition for such instability is the locally asymptotically stability of the homogeneous steady-state $E^{*}=(u^{*},v^{*})$ in the absence of diffusion, and it happens when $C_{1}=-(a_{11}+a_{22})>0$ and $C_{2}=a_{11}a_{22}-a_{12}a_{21}$ hold. Here, we apply heterogeneous perturbations around $E^{*}$ to obtain the criterion for such instability of the spatio-temporal model. For the case of one-dimensional diffusion, let us perturb the homogeneous steady state of the local system (\ref{eq:diff1}) around $(u^{*},v^{*})$ by 
\mathcenter
\begin{align*}
    \begin{pmatrix}
        u \\
        v
    \end{pmatrix}=\begin{pmatrix}
        u^{*} \\
        v^{*}
    \end{pmatrix}+\epsilon \begin{pmatrix}
        u_{1} \\
        v_{1}
    \end{pmatrix}e^{\lambda t+ikx}
\end{align*}
with $|\epsilon|\ll 1$ where $\lambda$ is the growth rate of perturbation and $k$ denotes the wave number. Substituting these values in system (\ref{eq:diff1}), the linearization takes the form:
\begin{equation} \label{eq:3.1}
\textbf{J}_{k}\begin{bmatrix}
u_{1}\\
v{1}
\end{bmatrix}\equiv \begin{bmatrix}
 a_{11}-d_{1}k^{2}-\lambda & a_{12} \\
 a_{21} & a_{22}-d_{2}k^{2}-\lambda 
\end{bmatrix}
\begin{bmatrix}
    u_{1} \\
    v_{1}
\end{bmatrix}
=\begin{bmatrix}
        0 \\
        0
    \end{bmatrix},
\end{equation}
where $a_{11},\ a_{12},\ a_{21}$ and $a_{22}$ are mentioned in Section \ref{subsec-2.2}. We are interested in finding the non-trivial solution of the system (\ref{eq:3.2}), so $\lambda$ must be a zero of $\det(\textbf{J}_{k})=0$, where $\textbf{J}_{k}$ is the coefficient matrix of the system (\ref{eq:3.2}). Now 
\begin{align*}
    \det(\textbf{J}_{k})=\lambda^{2}-B(k^{2})\lambda+C(k^{2})
\end{align*}
with $B(k^{2})=\mbox{tr}(\textbf{J}(E^{*}))-(d_{1}+d_{2})k^{2},\ C(k^{2})=\det(\textbf{J}(E^*))-(d_{2}a_{11}+d_{1}a_{22})k^{2}+d_{1}d_{2}k^{4}$. So, $\det(\textbf{J}_{k})=0$ gives 
\begin{align*}
    \lambda_{\pm}(k^{2})=\frac{B(k^{2})\pm\sqrt{(B(k^{2}))^{2}-4C(k^{2})}}{2}.
\end{align*}
Here, $B(k^{2})<0$ for all $k$ when the temporal model is locally asymptotically stable. So, the homogeneous solution remains stable under heterogeneous perturbation when $C(k^{2})>0$ for all $k$. If the inequality is violated for some $k \neq 0$, the system is unstable. In this case, $\displaystyle k^{2}_{min}=(d_{2}a_{11}+d_{1}a_{22})/2d_{1}d_{2}$ is the minimum value of $k^{2}$ for which $\displaystyle C(k^{2})$ will attain its minimum value, say $\displaystyle C(k^{2})_{min}$, such that
$$\displaystyle C(k^{2})_{min}=(a_{11}a_{22}-a_{12}a_{21})-\frac{(d_{2}a_{11}+d_{1}a_{22})^{2}}{4d_{1}d_{2}}.$$ 
This $k_{min}$ is the critical wave number for Turing instability. And the critical diffusion coefficient (Turing bifurcation threshold) $d_{2c}$ such that $C(k^{2})_{min}=0$ is given as 

\begin{equation} \label{eq:3.2}
d_{2c}=\frac{d_{1}(a_{11}a_{22}-2a_{12}a_{21})+d_{1}\sqrt{(a_{11}a_{22}-2a_{12}a_{21})^{2}-a_{11}^{2}a_{22}^{2}}}{a_{11}^{2}}.  
\end{equation}

The system will show stationary and non-stationary patterns for $d_{2}>d_{2c}$, but the coexisting equilibrium $(u^{*},v^{*})$ of the local model (\ref{eq:diff1}) remains stable under random heterogeneous perturbation when $d_{2}<d_{2c}$. Moreover, to ensure the positivity of $k^{2}=k^{2}_{\min}$ at the Turing bifurcation threshold, we need to have $d_{2}a_{11}+d_{1}a_{22}>0$, i.e., $d_{1}<d_{2}$ (as $a_{22}<0$) needs to be satisfied for the Turing instability conditions. Hence, the self-diffusion coefficient of the prey population is less than that of the predator population for the model (\ref{eq:diff1}). 

In addition, local stability analysis for the spatio-temporal model with two-dimensional diffusion will be the same as it is for the case of one-dimensional diffusion. In this case, the perturbation around $(u^{*}, v^{*})$ will take the following form: 
\mathcenter
\begin{align*}
    \begin{pmatrix}
        u \\
        v
    \end{pmatrix}=\begin{pmatrix}
        u^{*} \\
        v^{*}
    \end{pmatrix}+\epsilon \begin{pmatrix}
        u_{1} \\
        v_{1}
    \end{pmatrix}\exp{(\lambda t+i(k_{x}x+k_{y}y))},
\end{align*}
where $|\epsilon|\ll 1$ and $\lambda$ is the growth rate of perturbation. And, $\textbf{k}=(k_{x},k_{y})$ is the wave number vector along with the wave number $k=|\textbf{k}|$.

\section{Incorporation of nonlocal interaction} \label{sec:4}

Model (\ref{eq:diff1}) uses the assumption that the predator located at the space point $x$ makes an impact on the growth of prey at the same point. However, in reality, the fear of predators at a spatial location $x$ depends not only on the local appearance of the predator but also on the predator density in other nearby points, i.e., a prey located at space point $x$ can be scared by those predators who are located in some areas around this spatial point, and it can be obtained by convolution term
$$U(x,t):=(\Phi_{\sigma}*u)(x,t)=\int_{-\infty}^{\infty}\Phi_{\sigma}(x-y)u(y,t)dy.$$
Here, the kernel function $\Phi_{\sigma}(\cdot)$ describes the impact of fear on prey located at the spatial point by the predator located at some other spatial point. The first subscript $\sigma$ is the range of nonlocal interaction. We assume the kernel function $\Phi_{\sigma}$ to be non-negative with compact support in $\mathbb{R}$. Also, to preserve the same homogeneous steady-state solutions for both the local and nonlocal models, we assume that $\int_{-\infty}^{\infty}\Phi_{\sigma}(y)dy=1$. Now, the impact of fear on prey is limited by the number of predators they are facing or encountering. We can apply this limitation to each space point independently. This means that predators located at space point $y$ make an impact on prey at space point $x_{1}$ independently of its concentration at another point $x_{2}$. Considering the work of Furter and Grinfeld \cite{furter1989local} and using the aforementioned assumption, and the random movement of the population, we get the integro-differential equation model as:
\begin{align} \label{eq:loc1}
\frac{\partial u}{\partial t}&=d_{1}\frac{\partial^{2}u}{\partial x^{2}}+u(1-\Phi_{\sigma}*u)-\frac{\left(1-\frac{\omega v}{\phi+v}\right)uv}{m+\left(1-\frac{\omega v}{\phi+v}\right)u}, \nonumber \\
\frac{\partial v}{\partial t}&=d_{2}\frac{\partial^{2}v}{\partial x^{2}}+\frac{\xi\left(1-\frac{\omega v}{\phi+v}\right)uv}{m+\left(1-\frac{\omega v}{\phi+v}\right)u}-\delta v,
\end{align}
with non-negative initial and periodic boundary conditions.

\subsection{Linear stability analysis of the nonlocal model}

Both the local model (\ref{eq:diff1}) and the nonlocal model (\ref{eq:loc1}) show the same dynamics for homogeneous steady states. Let us consider $E^{*}=(u^{*},v^{*})$ as the coexisting homogeneous steady-state. Now, perturbing the system around $(u^{*},v^{*})$ by 
\begin{align*}
    \begin{pmatrix}
        u \\
        v
    \end{pmatrix}=\begin{pmatrix}
        u^{*} \\
        v^{*}
    \end{pmatrix}+\epsilon \begin{pmatrix}
        u_{1} \\
        v_{1}
    \end{pmatrix}e^{\lambda t+ikx}
\end{align*}
with $|\epsilon|\ll 1$ and substituting these values in system (\ref{eq:loc1}) the linearization takes the form:

\begin{equation} \label{eq:4.2}
\overline{\textbf{J}}_{k}\begin{bmatrix}
u_{1}\\
v{1}
\end{bmatrix}\equiv \begin{bmatrix}
 A_{11}-d_{1}k^{2}-\lambda & a_{12} \\
 a_{21} & a_{22}-d_{2}k^{2}-\lambda 
\end{bmatrix}
\begin{bmatrix}
    u_{1} \\
    v_{1}
\end{bmatrix}
=\begin{bmatrix}
        0 \\
        0
    \end{bmatrix},
\end{equation}
where $a_{12}, a_{21}$ and $a_{22}$ are mentioned in Section \ref{subsec-2.2} and
$$\displaystyle A_{11}=\frac{\{\phi+(1-\omega)v^{*}\}^{2}u^{*}v^{*}}{[m(\phi+v^{*})+\{\phi+(1-\omega)v^{*}\}u^{*}]^{2}}-u^{*}\frac{\sin k\sigma}{k\sigma}=a_{11}+u^{*}\left\{1-\left(\frac{\sin k\sigma}{k\sigma}\right)\right\}.$$
Now, we are interested in finding the non-trivial solution of the system (\ref{eq:4.2}), so $\lambda$ must be a zero of $\det(\overline{\textbf{J}}_{k})=0$, where $\overline{\textbf{J}}_{k}$ is the coefficient matrix of the system (\ref{eq:4.2}). Now 

\begin{equation*}
\det(\overline{\textbf{J}}_{k})=\lambda^{2}-\Gamma(k,d_{2},\sigma)\lambda+\Delta(k,d_{2},\sigma),
\end{equation*}
\mathleft
\begin{align*}
\mbox{where}\ \Gamma(k,d_{2},\sigma)&= (a_{11}+a_{22})-k^{2}(d_{1}+d_{2})+u^{*}\left(1-\frac{\sin k\sigma}{k\sigma}\right), 
\end{align*}
\begin{align*}
\mbox{and}\ \Delta(k,d_{2},\sigma)= d_{1}d_{2}k^{4}-\left[(a_{11}d_{2}+a_{22}d_{1})+d_{2}u^{*}\left(1-\frac{\sin k\sigma}{k\sigma}\right)\right]k^{2}+
\left[(a_{11}a_{22}-a_{21}a_{12})\right. \\
\left.+a_{22}u^{*}\left(1-\frac{\sin k\sigma}{k\sigma}\right)\right].
\end{align*}

So, $\det(\overline{\textbf{J}}_{k})=0$ gives
$\lambda(k^{2})=\left(\Gamma(k,d_{2},\sigma)\pm\sqrt{(\Gamma(k,d_{2},\sigma))^{2}-4\Delta(k,d_{2},\sigma)}\right)/2.$
For any fixed value of $d_{2}$ and $\sigma$, the homogeneous steady-state $(u^{*},v^{*})$ is stable when $\Gamma(k,d_{2},\sigma)<0$ and $\Delta(k,d_{2},\sigma)>0$ for all $k>0$. So, Turing instability occurs if for a fixed $d_{2}$, $\Gamma(k,d_{2},\sigma)<0$ for all $k$ and $\Delta(k,d_{2},\sigma)<0$ for some $k$. On the other hand, if $\Delta(k,d_{2},\sigma)>0$ for all $k>0$ and $\Gamma(k,d_{2},\sigma)>0$ for some $k$, then spatial-Hopf instability occurs.

\begin{enumerate}[(i)]
\item Turing instability: Let us assume that the equilibrium point $E^{*}$ is locally asymptotically stable for the temporal model. We need to fix the critical diffusion coefficient $d_{2c}^{N}$ and critical wave number $k_{c}^{N}$ so that $\Delta(k,d_{2},\sigma)=0$ for a unique $k=k_{c}^{N}$. Now, for all $d_{2}$, we have $\Delta(k,d_{2},\sigma)>0$ when $k$ is sufficiently small as well as a large quantity. Hence, $\Delta(k,d_{2},\sigma)=0$ holds for a unique $k$ when 
\mathcenter
\begin{equation} \label{eq:4.3}
 \Delta(k,d_{2},\sigma)=0\ \ \textrm{and}\ \ \frac{\partial \Delta(k,d_{2},\sigma)}{\partial k}=0  
\end{equation}
which gives us 
\begin{subequations}\label{eq:4.4}
\begin{align} 
&d_{2}=\frac{d_{1}a_{22}k^{2}-\left[(a_{11}a_{22}-a_{12}a_{21})+a_{22}u^{*}\left(1-\frac{\sin k\sigma}{k\sigma}\right)\right]}{d_{1}k^{4}-k^{2}\left\{a_{11}+u^{*}\left(1-\frac{\sin k\sigma}{k\sigma}\right)\right\}}, \label{eq:4.4a} \\
&4d_{1}d_{2}k^{3}+\frac{d_{2}u^{*}}{k\sigma}\left(\sigma \cos k\sigma-\frac{\sigma \sin k\sigma}{k\sigma}\right)k^{2}-2\left[(d_{2}a_{11}+d_{1}a_{22})+d_{2}u^{*}\left(1-\frac{\sin k\sigma}{k\sigma}\right)\right]k \nonumber \\
&-\frac{a_{22}u^{*}}{k\sigma}\left(\sigma \cos k\sigma-\frac{\sigma \sin k\sigma}{k\sigma}\right)=0. \label{eq:4.4b}
\end{align}
\end{subequations}

From (\ref{eq:4.4}), eliminating $d_{2}$ leads to the following transcendental equation

\begin{equation} \label{eq:4.5}
\begin{aligned}
2a_{22}\left[u^{*}\left(1-\frac{\sin k\sigma}{k\sigma}\right)-d_{1}k^{2}\right]^{2}-2(2d_{1}k^{2}-a_{11})(a_{11}a_{22}-a_{12}a_{21})\\
+2u^{*}(2a_{11}a_{22}-a_{12}a_{21})\left(1-\frac{\sin k\sigma}{k\sigma}\right)+\frac{a_{12}a_{21}u^{*}}{\sigma}\left(\cos k\sigma-\frac{\sin k\sigma}{k\sigma}\right)=0.
\end{aligned}
\end{equation}
Solving the equation we get a critical wave number $k_{c}^{N}$, and substituting the value in (\ref{eq:4.4a}) we get $d_{2c}^{N}$ as the critical diffusion coefficient, provided $\Gamma(k,d_{2c}^{N},\sigma)<0$ for all $k>0$. Now, being a transcendental equation, (\ref{eq:4.5}) may have more than one root. In that case, we have to choose that $k_{c}^{N}$ for which $d_{2c}^{N}$ will be a positive minimum number. Here $d_{2}>d_{2c}^{N}$ leads to $\Delta(k,d_{2},\sigma)<0$ for a range of values of $k$ and a fixed value of $\sigma$, so Turing instability occurs for $d_{2}>d_{2c}^{N}$. We consider $\Delta_{\min}$ as the minimum of $\Delta$ for $k>0$.

\item Spatial-Hopf instability: In the temporal-Hopf stable region, $\Gamma(k,d_{2},\sigma)<0$ when $k$ is sufficiently small as well as a large quantity. For spatial-Hopf instability condition, we need to find a critical diffusion coefficient $d_{2c}^{H}$ for which $\Gamma(k,d_{2},\sigma)=0$ for a unique $k=k_{c}^{H}$. So, $\Gamma(k,d_{2},\sigma)=0$ holds for a unique $k$ when 
\mathcenter
\begin{equation} \label{eq:4.6}
 \Gamma(k,d_{2},\sigma)=0\ \ \textrm{and}\ \ \frac{\partial \Gamma(k,d_{2},\sigma)}{\partial k}=0  
\end{equation}
hold, which gives 
\begin{subequations}\label{eq:4.7}
\begin{align} 
d_{2}&=\frac{a_{11}+a_{22}+u^{*}\left(1-\frac{\sin k\sigma}{k\sigma}\right)}{k^{2}}-d_{1} \label{eq:4.7a} \\
\mbox{and}\ \ & 2k(d_{1}+d_{2})+\frac{u^{*}}{k}\left(\cos k\sigma-\frac{\sin k\sigma}{k\sigma}\right)=0. \label{eq:4.7b}
\end{align}
\end{subequations}
From (\ref{eq:4.7}), eliminating $d_{1}$ leads to the following transcendental equation
\begin{equation} \label{eq:4.9}
\begin{aligned}
2\left[a_{11}+a_{22}+u^{*}\left(1-\frac{\sin k\sigma}{k\sigma}\right)\right]+u^{*}\left(\cos k\sigma-\frac{\sin k\sigma}{k\sigma}\right)=0.
\end{aligned}
\end{equation}
Solving \eqref{eq:4.9} we get a critical wave number $k_{c}^{H}$ which has to be substituted in (\ref{eq:4.7a}) in order to get spatial-Hopf bifurcation threshold $d_{2c}^{H}$, provided $\Delta(k,d_{2c}^{H},\sigma)>0$ for all $k>0$. Equation (\ref{eq:4.9}) may possess multiple zeros, out of which we have to choose that $k_{c}^{H}$ for which $d_{2c}^{H}$ will be minimum. Here $d_{2}<d_{2c}^{H}$ leads to $\Gamma(k,d_{2},\sigma)>0$ for a range of values of $k$ and a fixed value of $\sigma$ and we choose $\Gamma_{\max}$ as the maximum of $\Gamma$ for $k>0$.
\end{enumerate}

It is evident that $\Gamma_{\max}=0$ and $\Delta_{\min}>0$ correspond to spatial Hopf bifurcation point.
Also, $\Gamma_{\max}<0$ and $\Delta_{\min}=0$ corresponds to Turing bifurcation point. The curve with $\Delta_{\min}=0$ is commonly referred to as the Turing curve, whereas the segment of the curve with $\Gamma_{\max}<0$ is known as the Turing bifurcation curve. Depending on $\sigma$, we have two possibilities: either the curves $\Gamma_{\max}=0$ and $\Delta_{\min}=0$ intersect or do not
intersect in the $\omega$-$d_{2}$ plane. We shall numerically validate these two cases in the following section.

\section{Numerical Results} \label{sec:5}
The numerical simulation is usually performed to validate the analytical results. Here, we have used MATLAB to illustrate the dynamical scenarios of the proposed system in the absence of and in the presence of species diffusion. Unless mentioned otherwise, we fix some of the parameters used in the model, which are given in Table \ref{Table:1}. 

\begin{table}[!ht]
\centering
\begin{tabular}{|c|c|c|c|c|c|}
\hline
Parameters & $\xi$ & $\phi$ & $\delta$ & $\omega$ & $m$ \\
 \hline
Value & $0.6$ & $0.2$ & $0.43$ & $0.85$ & $0.113$ \\
\hline
\end{tabular}
\caption{Parameter values used in the numerical simulations} \label{Table:1}
\end{table}

\begin{figure}[htb!]
    \centering
     \centering
    \begin{subfigure}[t]{0.5\textwidth}
        \centering
        \includegraphics[width=7.5cm,height=5.6cm]{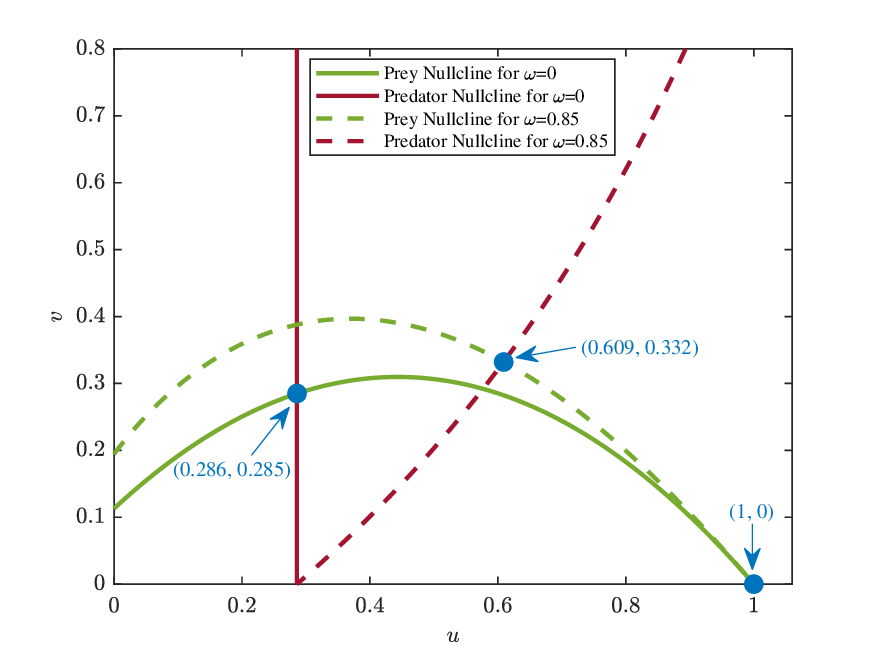}
        \caption{}\label{fig:1a}
    \end{subfigure}%
    ~ 
    \begin{subfigure}[t]{0.5\textwidth}
        \centering
        \includegraphics[width=7.7cm,height=6cm]{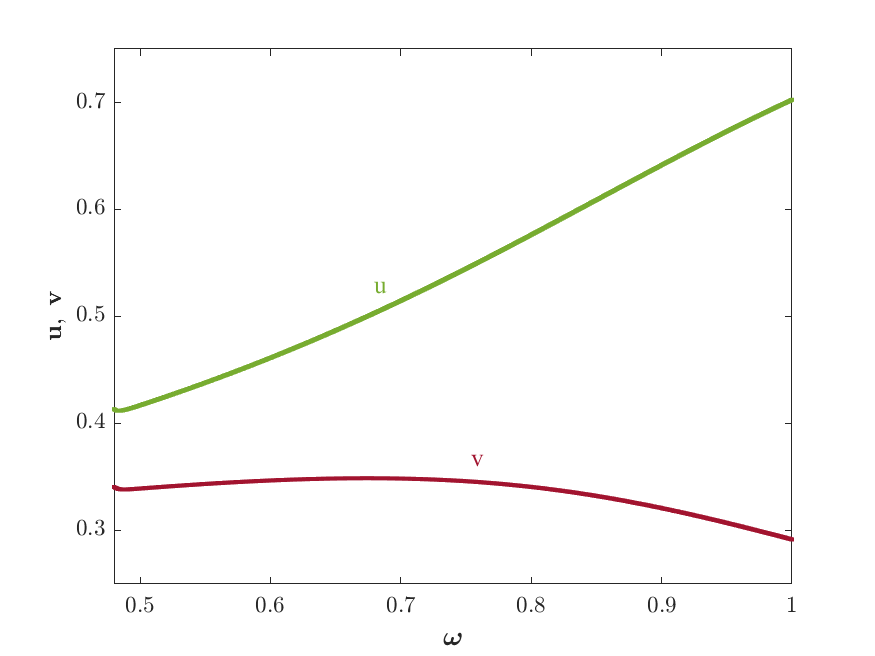}
        \caption{}\label{fig:1b}
    \end{subfigure}
    \caption{(a) The nontrivial prey are predator nullcline is represented when $\omega=0$ and $\omega\neq 0$. The nontrivial prey nullcline intersects the horizontal axis at $(1, 0)$. (b) Trajectory profiles of prey $(u)$ and predator $(v)$ species in presence of inducible defense. Parameter values are mentioned in Section \ref{sec:5}.} \label{fig:1}
\end{figure}

In Fig. \ref{fig:1a}, we have plotted the non-trivial prey and predator nullclines, which reflects how the incorporation of inducible defence makes changes in species count. It is observed that the prey species significantly increases with the increased defense level leading to a decline in the predator population [see Fig. \ref{fig:1}].

\begin{figure}[htb!]
    \centering
    \begin{subfigure}{.32\textwidth}
        \centering
        \includegraphics[width=6cm]{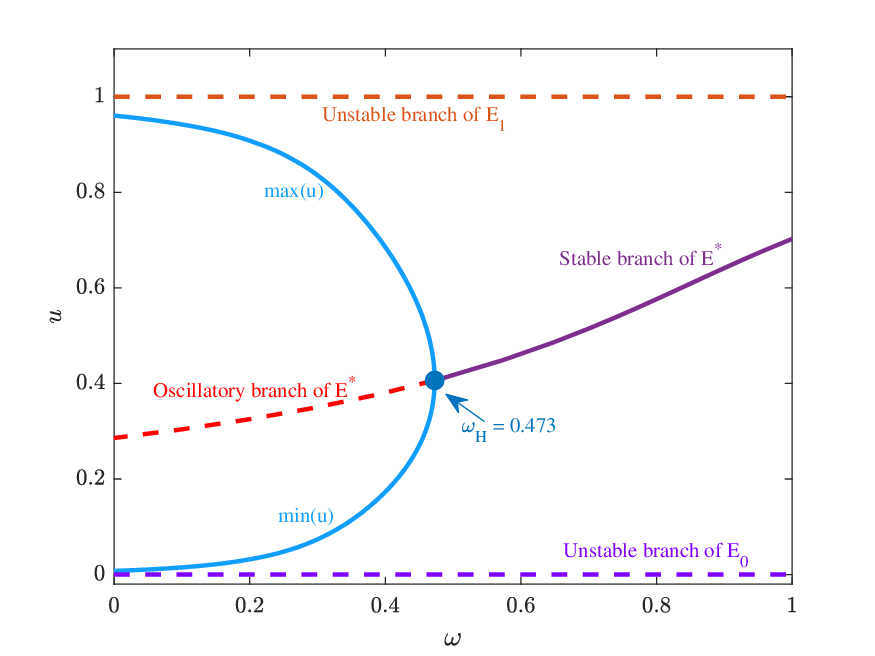}
        \caption{}\label{fig:2a}
    \end{subfigure}
    \begin{subfigure}{.32\textwidth}
        \centering
        \includegraphics[width=6cm]{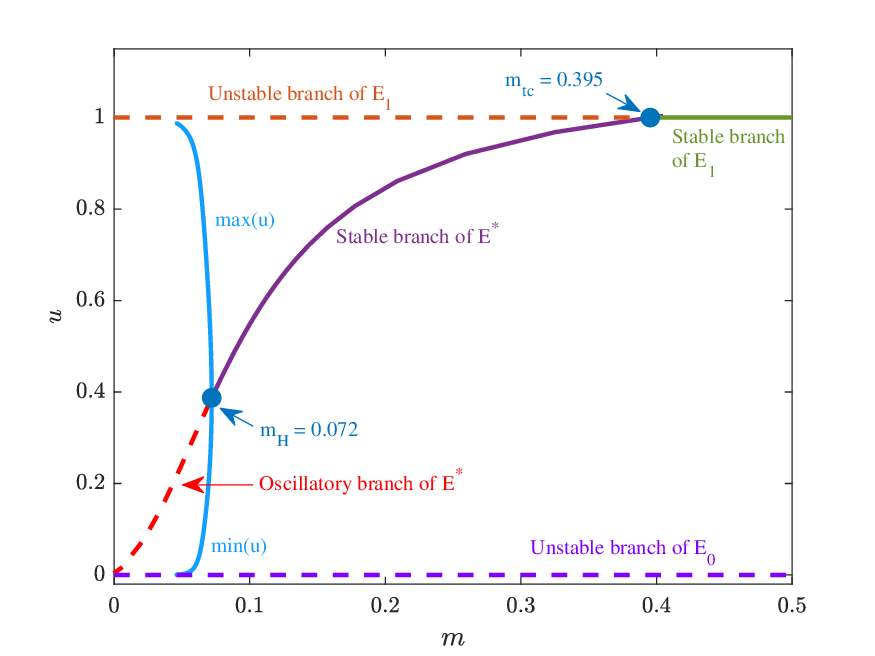}
        \caption{}\label{fig:2b}
    \end{subfigure}
    \begin{subfigure}{.32\textwidth}
        \centering
        \includegraphics[width=6cm]{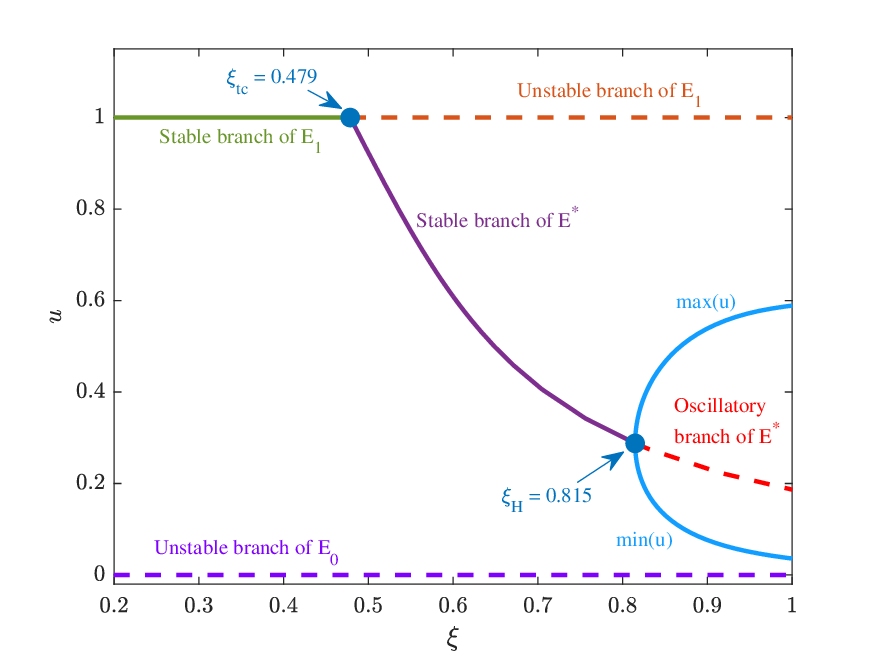}
        \caption{}\label{fig:2c}
    \end{subfigure}
    \caption{Stabilities of the equilibrium points for the temporal model (\ref{eq:det1}) for the bifurcation parameters (a) $\omega$, (b) $m$ and (c) $\xi$.} \label{fig:2}
\end{figure}

Figure \ref{fig:2} shows how the parameters $\omega$ and $m$ change the stability of the temporal system (\ref{eq:det1}). Particularly, Fig. \ref{fig:2a} delineates the stabilizing effect of the inducible defence as increasing defence level leads to a stable coexisting state. It shows oscillatory behaviour for lower defence levels, and the amplitudes of the oscillations decrease with an increase in $\omega$. In addition, both species' densities settle to a stable interior state when $\omega$ exceeds a certain threshold value, which occurs through a supercritical Hopf bifurcation around $E^{*}$ at $\omega_{H}=0.473$ for the chosen parametric values. On the other hand, the parameter $m$ also plays an important role in the system dynamics. The predator-free equilibrium $(E_{1})$ switches its dynamics from an unstable to stable state when $m$ increases and the system exhibits a transcritical bifurcation around $E_{1}$ at $m_{tc}=0.395$ [see Fig. \ref{fig:2b}]. The coexisting equilibrium also shows an oscillating nature while $m$ holds a minimal value, but these oscillations do not last long as slightly higher values of $m$, which leads the species to a stable situation. This stability switching occurs through a supercritical Hopf bifurcation around $E^{*}$ at $m_{H}=0.072$, and the species settles down to a stable interior state for $m>m_{H}$. Further, an increase in $m$ indicates a proliferation in prey species, which causes predator declination, and ultimately, it coincides with the predator-free state through the transcritical bifurcation at $m=m_{tc}$. Moreover, Fig. \ref{fig:2c} portrays how the changes in biomass conversion rate $(\xi)$ make an impact on the dynamics of the system. The predator-free state $(E_{1})$ switches its dynamics from a stable to unstable situation when for increasing value of $\xi$ and the system undergoes a transcritical bifurcation around $E_{1}$ at $\xi_{tc}=0.479$. On the other hand, the coexisting equilibrium also shows oscillation when $\xi$ holds a larger value, but the amplitude reduces with the decrement of $\xi$, which leads the species to coexist in a stable situation. This stability switching occurs through a supercritical Hopf bifurcation around $E^{*}$ at $\xi_{H}=0.815$, and the species settles down to a stable interior state for $\xi<\xi_{H}$. Further decrease in $\xi$ indicates a diminution in predator species, and ultimately, it coincides with the predator-free state through the transcritical bifurcation at $\xi=\xi_{tc}$. Altogether, the system parameters $\omega,\ m$, and $\xi$ have a noteworthy contribution to the system's dynamics.


\begin{figure}[htb!]
    \centering
        \includegraphics[height=6.5cm,width=10cm]{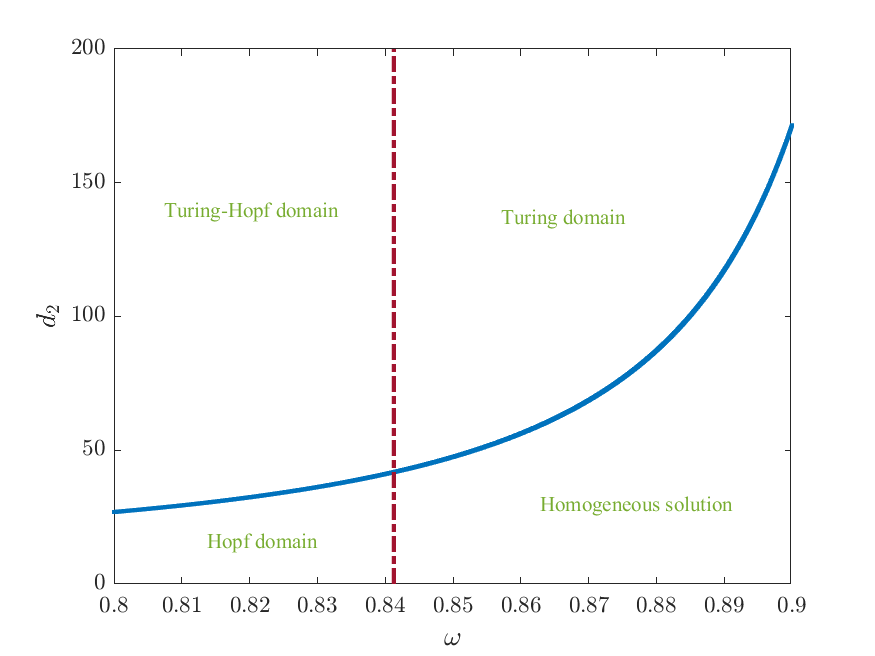}
    \caption{Temporal-Hopf ({\color{darkmar}\protect\dashdotrule}), Turing bifurcation ({\color{fadblu}\solidrule}) curves for the local model in the $\omega$-$d_{2}$ plane.} \label{fig:3}
\end{figure}

The temporal model is studied further in the presence of species diffusion. To describe the Turing and non-Turing patterns for the system \eqref{eq:diff1}, it is assumed that the species move in a one-dimensional smooth spatial domain $\Omega=[-L, L]$, where $L=200$ with non-negative initial and periodic boundary conditions. We have chosen $dx=0.5$ and $dt=0.0005$. For simulations, we have fixed the parameter values $\xi=0.95$ and $m=0.128$, and the prey diffusion coefficient $d_{1}$ is chosen as $0.8$. In this case, the Hopf threshold for the defence level becomes $\omega_{H}=0.841$, and the system becomes stable for $\omega>\omega_{H}$ from the oscillatory state. The analytical results state that Turing instability occurs in the system when $d_{2}>d_{2c}$. Figure \ref{fig:3} depicts the interdependence of Turing bifurcation between the inducible defence and species diffusion. To observe the spatio-temporal dynamics for the parameters lie in different domains created by the Turing and Hopf curves, a heterogeneous perturbation is given around the coexisting homogeneous steady-state as the initial conditions. We choose small amplitude random perturbations given by $u(x,0)=u^{*}+\epsilon\eta(x)$ and $v(x,0)=v^{*}+\epsilon\psi(x)$ with $\epsilon=10^{-4}$ and $\eta$ and $\psi$ are Gaussian white noise $\sigma$-correlated in space.

\begin{figure}[htb!]
    \centering
    \begin{subfigure}{.32\textwidth}
    \centering
    \includegraphics[width=\linewidth]{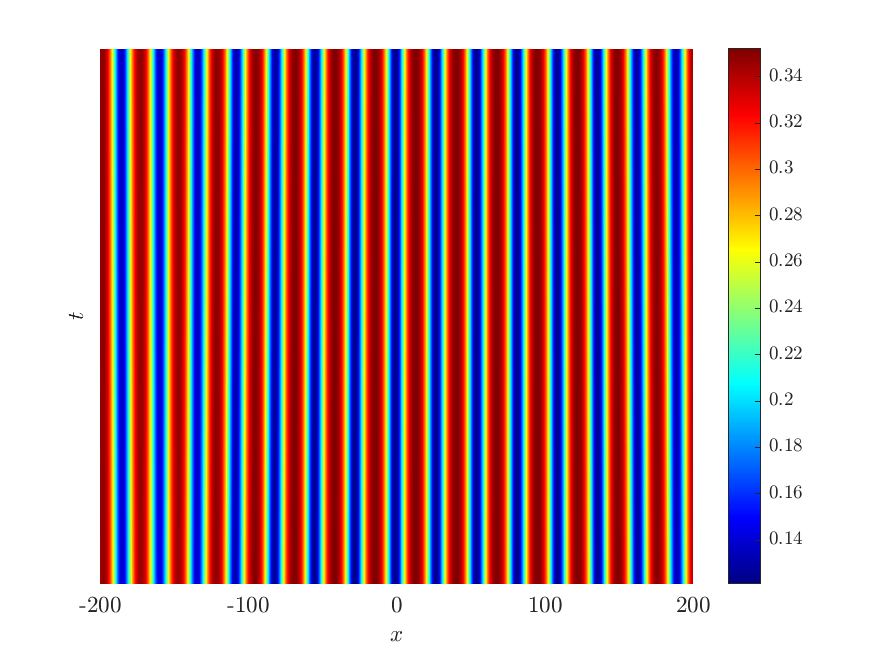}  
    \caption{}\label{fig:4a}
\end{subfigure}
\begin{subfigure}{.32\textwidth}
    \centering
    \includegraphics[width=\linewidth]{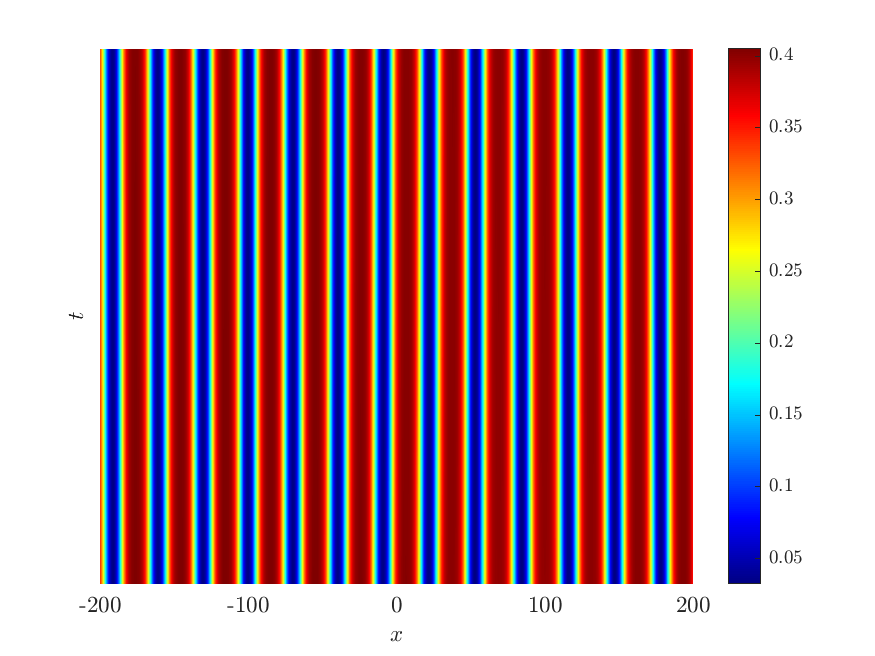}  
    \caption{}\label{fig:4b}
\end{subfigure}
\begin{subfigure}{.32\textwidth}
    \centering
    \includegraphics[width=\linewidth]{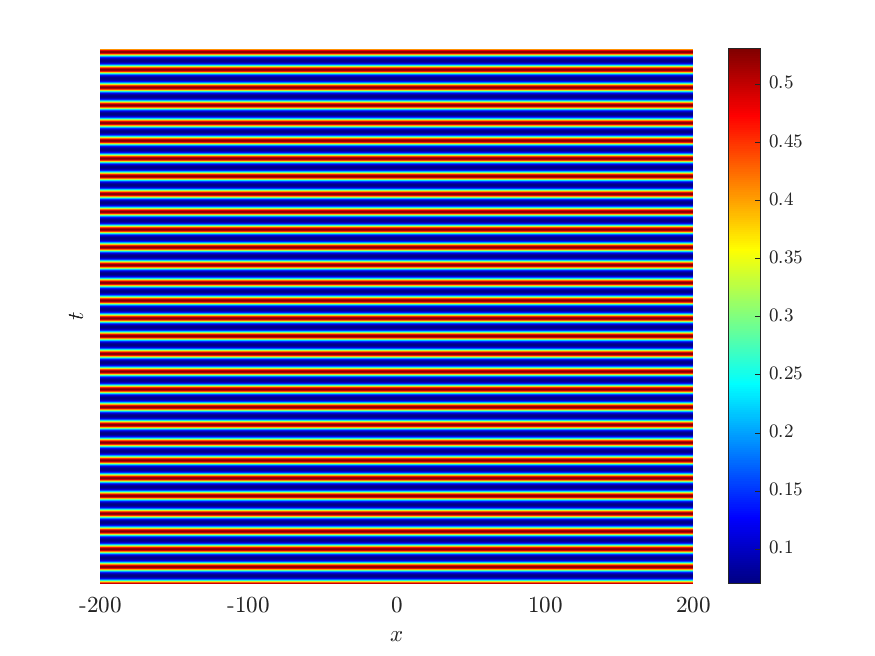}  
    \caption{}\label{fig:4c}
\end{subfigure}
    \caption{Contour plots of $u$ of the local model \eqref{eq:diff1} for $t\in [2000,2550]$ when $(\omega, d_{2})$ chosen as (a) $(0.85, 60)$: Turing domain, (b) $(0.8, 80)$: Turing-Hopf domain, and (c) $(0.8, 10)$: Hopf domain.} \label{fig:4}
\end{figure}

\begin{figure}
\centering
\begin{subfigure}{.32\textwidth}
    \centering
    \includegraphics[width=\linewidth]{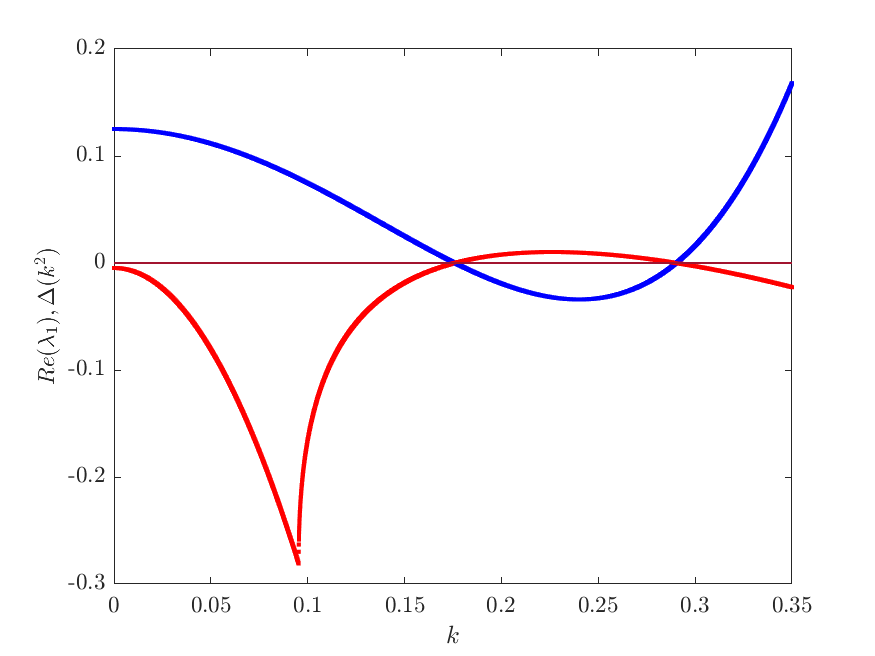}  
    \caption{}\label{fig:5a}
\end{subfigure}
\begin{subfigure}{.32\textwidth}
    \centering
    \includegraphics[width=\linewidth]{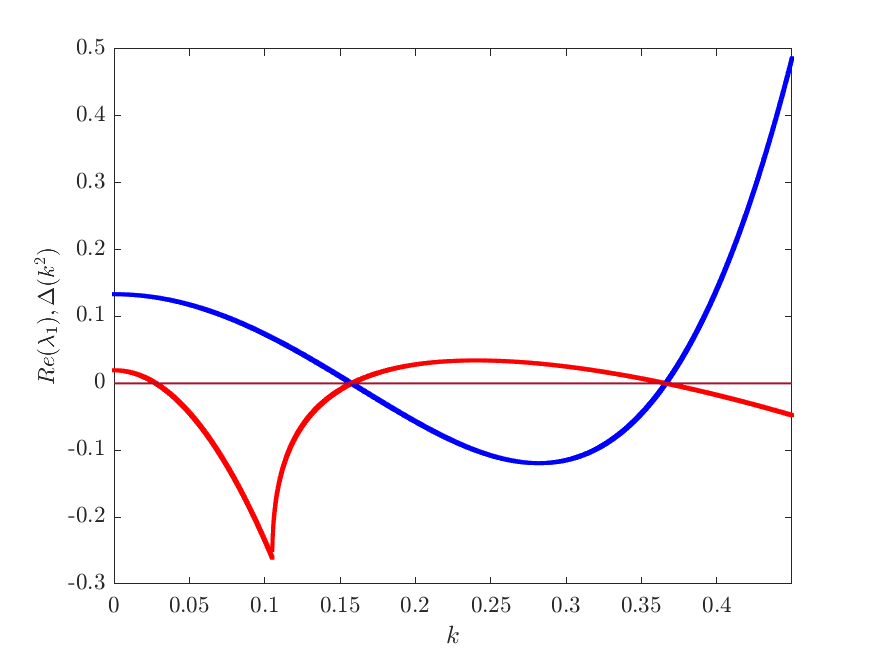}  
    \caption{}\label{fig:5b}
\end{subfigure}
\begin{subfigure}{.32\textwidth}
    \centering
    \includegraphics[width=\linewidth]{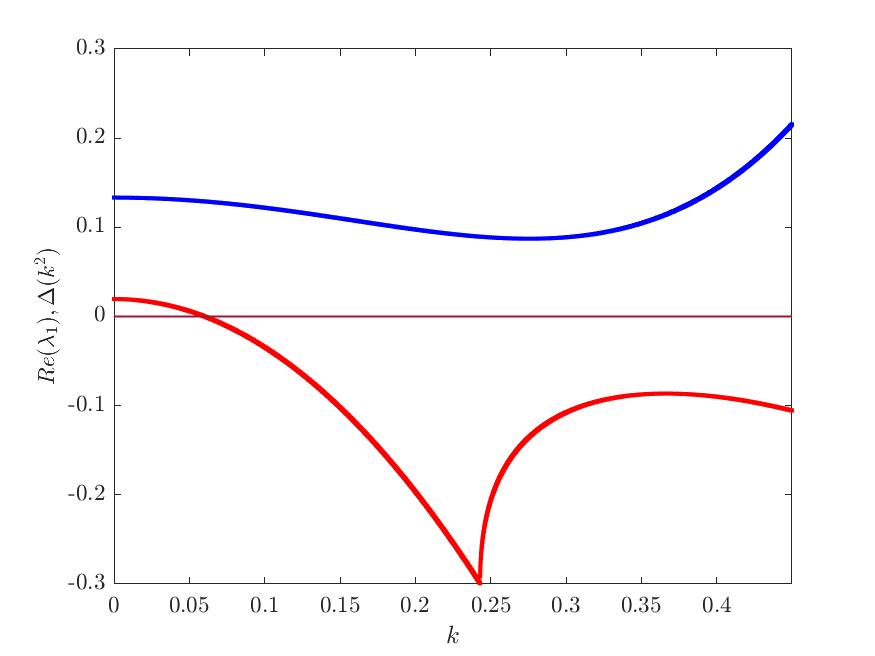}  
    \caption{}\label{fig:5c}
\end{subfigure}
\caption{Plots of $\max(\mbox{Re}(\lambda))$ ({\color{red}\rule[0.5mm]{6mm}{2pt}}) and $\Delta(k^{2})$ ({\color{blue} \rule[0.5mm]{6mm}{2pt}}) with respect to $k$ for different values of $(\omega, d_{2})$: (a) $(0.85,60)$ in Turing domain, (b) $(0.8,80)$ in Turing-Hopf domain, and (c) $(0.8,10)$ in Hopf domain.
} \label{fig:5}
\end{figure}

The dynamical behaviour of the proposed spatio-temporal model is explored in Fig. \ref{fig:4}. For $\omega=0.85$, the temporal model has the feasible interior equilibrium point $E^{*}=(0.247,0.411)$, which is stable. And, equation (\ref{eq:3.2}) gives the Turing bifurcation threshold as $d_{2c} = 47.532$ for $d_{1}=0.8$. In Fig. \ref{fig:4}, the spatio-temporal patterns have been portrayed when the species disperse in a one-dimensional domain, and $(\omega, d_{2})$ are chosen from Turing, Turing-Hopf, and Hopf domains. Fig. \ref{fig:4a} depicts the stationary Turing patterns for the periodic boundary condition when $(\omega, d_{2})$ lies in the Turing domain. Moreover, non-homogeneous stationary patterns are observed in the Turing-Hopf domain [see Fig. \ref{fig:4b}], while oscillatory solutions can be found in the Hopf domain [see Fig. \ref{fig:4c}]. To mathematically justify the occurrence of Fig. \ref{fig:5}, the real part of an eigenvalue and their corresponding determinant are plotted in Fig. \ref{fig:5} for each of the mentioned cases. Not to mention that if we choose $(\omega, d_{2})$ from the domain containing homogeneous solution, then $Re(\lambda)$ will always be negative for all $k>0$.


These findings indicate that an elevated defence level in prey species increases the predator diffusion threshold $d_{2}$ for the emergence of Turing patterns, thereby constricting the parametric region conducive to patch formation. So, a higher level of defence in prey contracts the chances of patch formation. Inducible defences often make prey harder to catch or less appealing to predators. A reduced Turing domain implies that these defences confine the places where predators may readily capture prey, potentially enhancing prey survival rates. Predator-prey interactions may become more localized when the Turing domain is shrunk. Turing patterns help to maintain ecosystem diversity and stability by forming niches and sustaining a wide range of species. A smaller area might diminish spatial complexity, perhaps resulting in a more homogeneous environment with distinct consequences for biodiversity. 

\begin{figure}[htb!]
    \centering
    \begin{subfigure}[t]{0.32\textwidth}
        \centering
        \includegraphics[width=\linewidth]{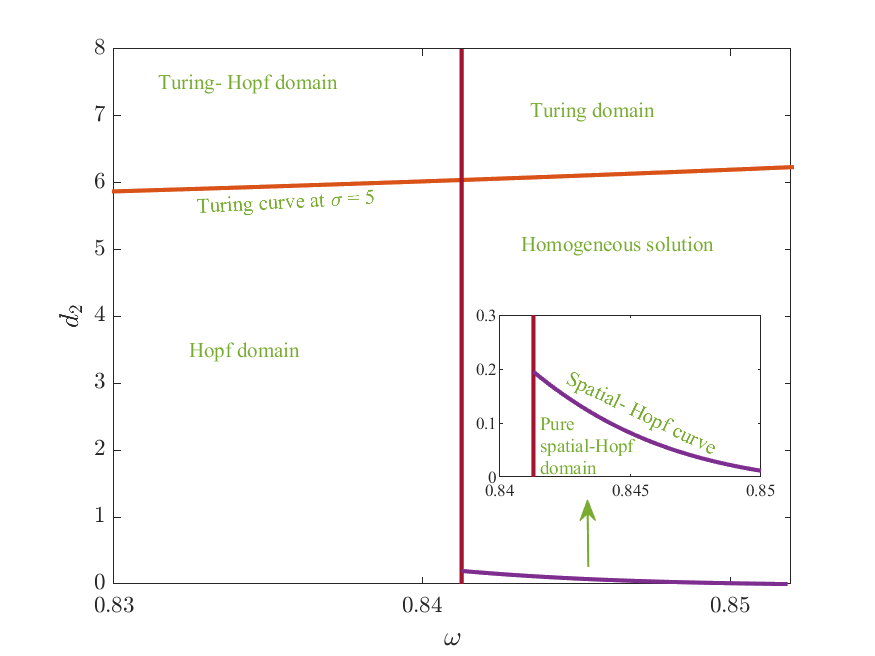}
        \caption{}\label{fig:6b}
    \end{subfigure}
    \begin{subfigure}[t]{0.32\textwidth}
        \centering
        \includegraphics[width=\linewidth]{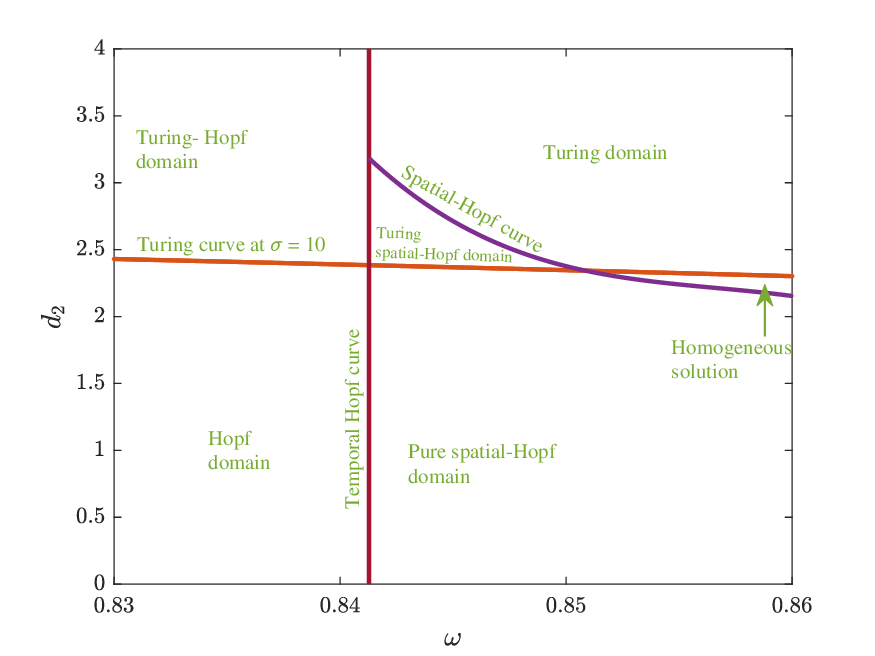}
        \caption{}\label{fig:6c}
    \end{subfigure}%
    \begin{subfigure}[t]{0.32\textwidth}
        \centering
        \includegraphics[width=\linewidth]{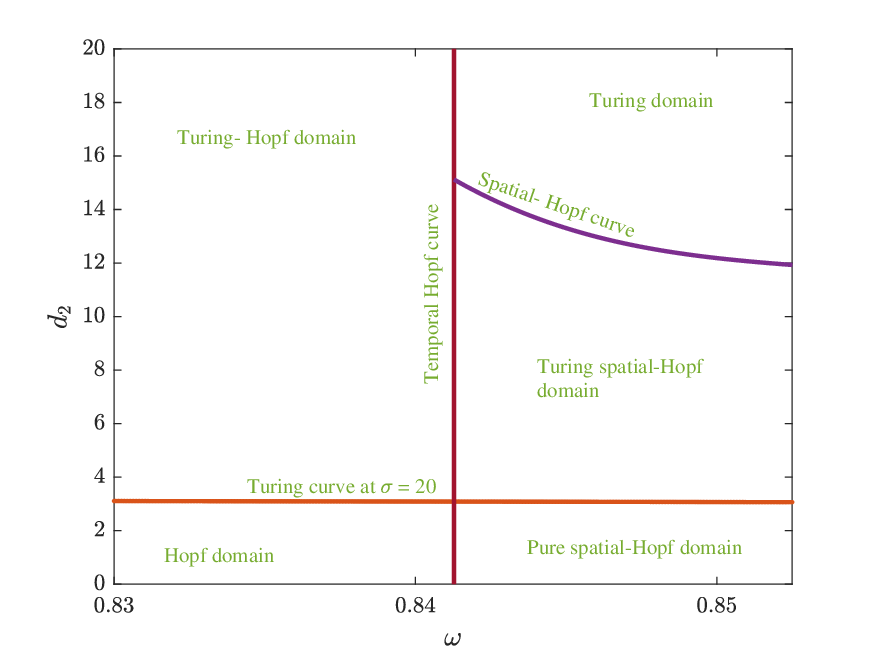}
        \caption{}\label{fig:6d}
    \end{subfigure}
    \caption{Bifurcation curves for the nonlocal model with top-hat kernel for different values of $\sigma$: (a) $\sigma=5$, (b) $\sigma=10$, and (c) $\sigma=20.$ The temporal Hopf, Turing, and spatial-Hopf curves are shown in each figure. } \label{fig:6}
\end{figure}

We investigate the behaviour of the nonlocal model (\ref{eq:loc1}) for various values of the range of nonlocal interactions $\sigma$. It is also noted that the nonlocal model \eqref{eq:loc1} turns into a local model \eqref{eq:diff1} if the range of nonlocal interaction $\sigma$ tends to $0$. In addition, the range of nonlocal interaction $\sigma$ does not affect the temporal Hopf bifurcation curve; however, it affects the Turing bifurcation curve. Furthermore, the range of nonlocal interaction produces spatial-Hopf bifurcation, which does not occur for the local model. Now, we discuss the qualitative and quantitative behaviour of the solutions of the nonlocal model in the one-dimensional spatial domain $L=[-500, 500]$ with $dx=0.25$ and $dt=0.001$. 

The findings illustrated in Fig. \ref{fig:3} demonstrate a contraction in the Turing domain as prey defence increases, thereby reducing the likelihood of pattern formation. On the other hand, Fig. \ref{fig:6} reveals that introducing nonlocal interactions significantly broadens this domain. However, this incorporation of nonlocality also introduces the spatial-Hopf curve. Specifically, in Fig. \ref{fig:6b}, a modest value of nonlocal range $(\sigma=5)$ ensures that the spatial-Hopf curve remains consistently below the Turing curve, even if the prey's inducible defence intensifies. As the nonlocal interaction range increases, this spatial-Hopf curve shifts upwards with the level of prey defence, leading to a contraction of the Turing domain while expanding the Turing spatial-Hopf domain. Notably, the spatial-Hopf curve exhibits a monotonic decrease with increasing inducible defence levels. From an ecological perspective, these dynamics in a predator-prey model suggest that an increase in prey defence mechanism not only suppresses the conditions for Turing patterns but also influences the stability landscape. Incorporating nonlocal interactions adds a new layer of complexity; it not only broadens the conditions for spatial pattern formation but also pushes the system toward regions where oscillatory patterns (spatial-Hopf) may emerge. This interaction between localized defence mechanisms and the range of nonlocal interactions highlights how a spatial organization in ecological interactions can give rise to oscillatory or patterned distributions within prey-predator populations.

\begin{figure}[htb!]
    \centering
    \begin{subfigure}[t]{0.5\textwidth}
        \centering
        \includegraphics[width=7cm]{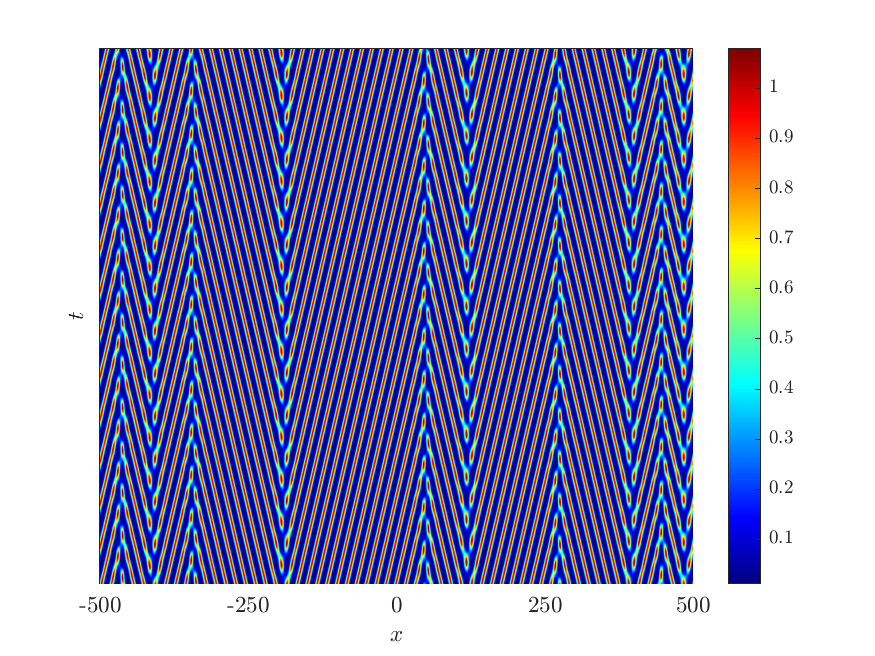}
        \caption{}\label{fig:11a}
    \end{subfigure}%
    ~
    \begin{subfigure}[t]{0.5\textwidth}
        \centering
        \includegraphics[width=7cm]{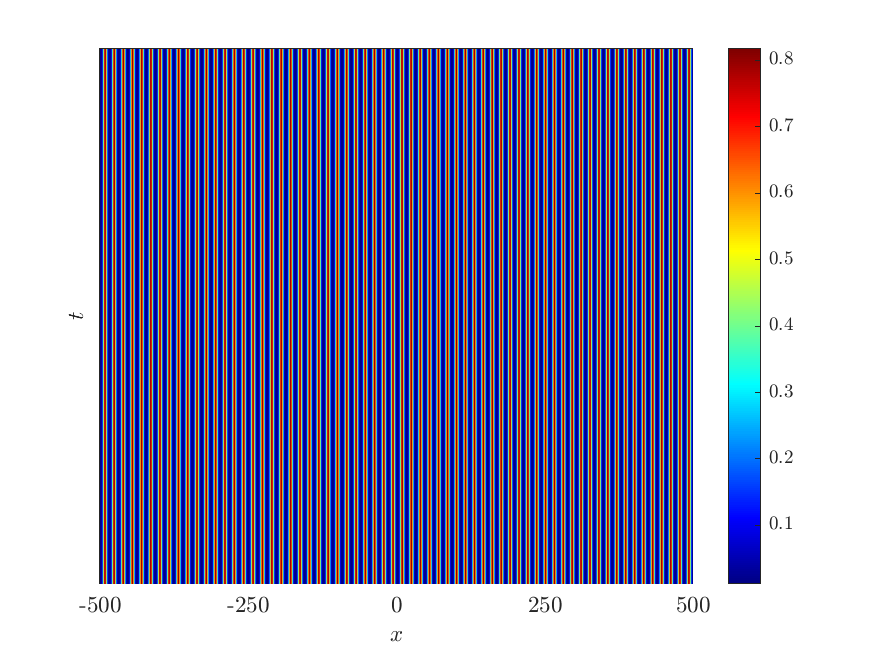}
        \caption{}\label{fig:11b}
    \end{subfigure}
    \caption{Formation of different types of patterns for $t\in [1000,1300]$ with $\sigma=5$ when $\omega =0.842$ is chosen from Hopf stable domain and (a) $d_{2}=0.1$ (pure spatial-Hopf domain); and (b) $d_{2}=8$ (Turing domain).} \label{fig:11}
\end{figure}


We first fix $\sigma=5$, and the domains separated by the Turing and spatial-Hopf curves are shown in Fig. \ref{fig:6b}. Here, the curve on which $\Gamma_{\max}=0$ and $\Delta_{\min}>0$ hold is known as the spatial-Hopf bifurcation curve. In the considered parametric region, the $\omega$-$d_{2}$ plane is divided into five sub-regions. We have $d_{2c}^{H}<d_{2c}^{N}$ when $\omega$ lies in the Hopf stable region. In the pure spatial-Hopf domain, $\Gamma_{\max}>0$ and $\Delta_{\min}>0$ hold. Hence, $\Gamma>0$ for a range of value of $k$ and $\Delta_{\min}>0$ for a range of value of $k$ as well in this region. These two ranges of values of $k$ may or may not overlap. Hence, we get oscillatory solutions or non-homogeneous steady patterns depending on the signs of $\Gamma$ and $\Delta$ for $k>0$ [see Fig. \ref{fig:11a}].  And, in the domain of homogeneous solution, we have $\Gamma_{\max}<0$ and $\Delta_{\min}>0$. Stationary non-homogeneous patterns are observed in the Turing region where $\Delta_{\min}<0$ and $\Gamma_{\max}<0$ hold [see Fig. \ref{fig:11b}]. The Hopf unstable and Turing-Hopf regions are also depicted where we observe oscillatory solution and non-homogeneous stationary pattern as well as oscillatory solution, respectively. Numerical simulations in Fig. \ref{fig:11} show a two-periodic non-homogeneous solution in the pure spatial-Hopf domain and a stationary Turing pattern in the Turing domain. Furthermore, introducing nonlocal interaction broadens the Turing domain by raising the probability of stationary patterns in the population.

\begin{figure}[htb!]
    \centering
    \begin{subfigure}[t]{0.3\textwidth}
        \centering
        \includegraphics[width=6.2cm]{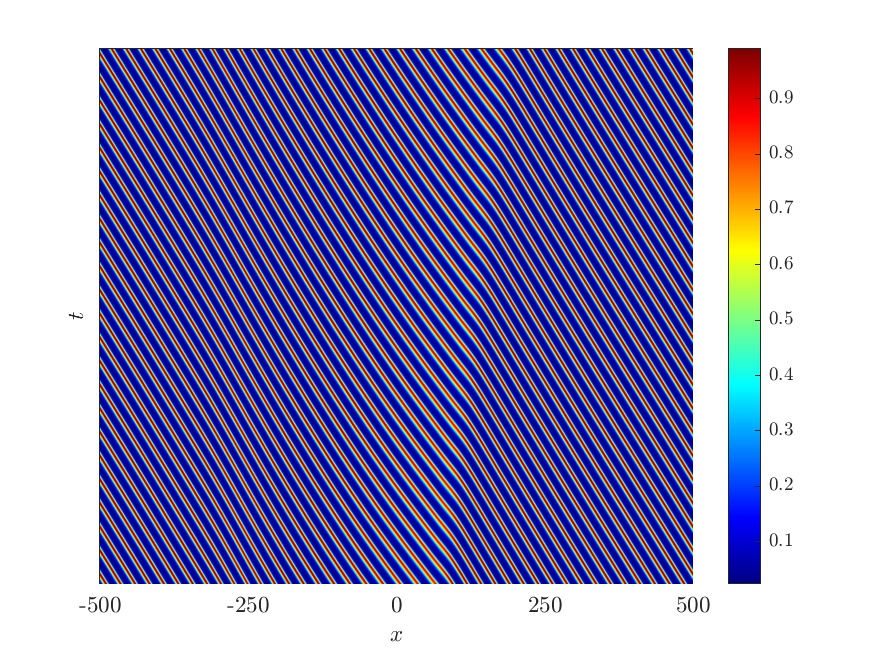}
        \caption{}\label{fig:8a}
    \end{subfigure}%
    ~
    \begin{subfigure}[t]{0.3\textwidth}
        \centering
        \includegraphics[width=6.2cm]{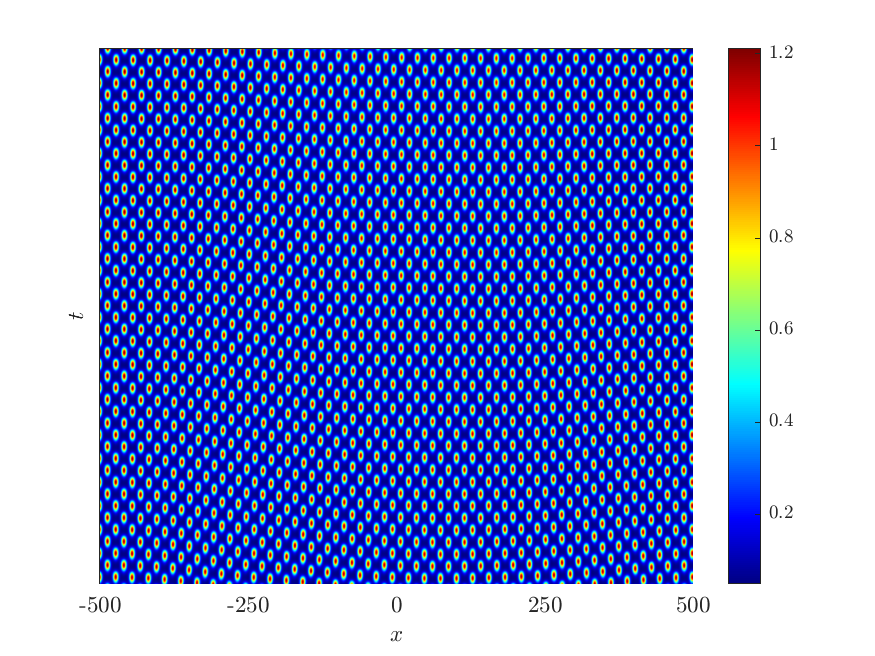}
        \caption{}\label{fig:8b}
    \end{subfigure}
    ~
    \begin{subfigure}[t]{0.3\textwidth}
        \centering
        \includegraphics[width=6.2cm]{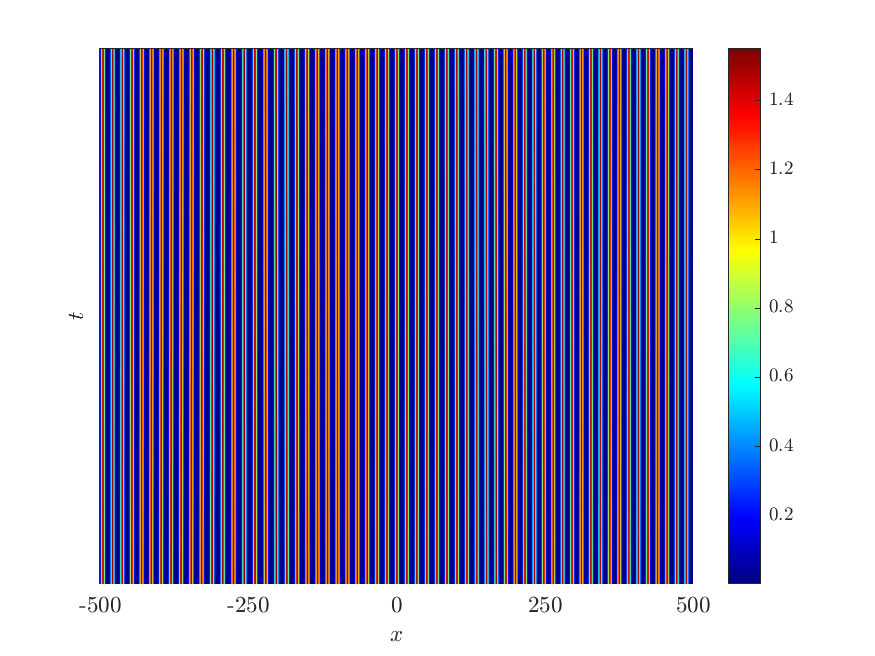}
        \caption{}\label{fig:8c}
    \end{subfigure}
    \caption{Color plot of $u$ for $t\in[2500,3000]$ with $\sigma=10$ when $\omega$ is chosen from Hopf stable domain and $d_{2}$ lies below or above the spatial- Hopf curve. (a) $d_{2}<d_{2c}^{N}$ (lies below the Turing curve), (b) $d_{2c}^{N}<d_{2}<d_{2c}^{H}$ (lies above the Turing curve but below the spatial-Hopf curve), (c) $d_{2}>d_{2c}^{H}$ (lies above the spatial-Hopf curve).} \label{fig:8}
\end{figure}

\begin{figure}[htb!]
    \centering
    \begin{subfigure}[t]{0.3\textwidth}
        \centering
        \includegraphics[width=6.2cm]{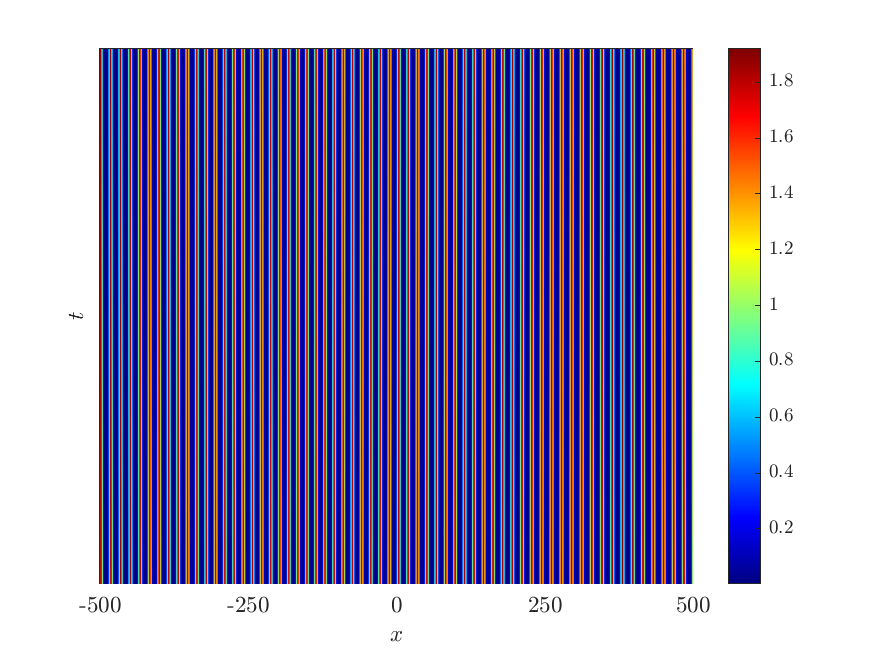}
        \caption{}\label{fig:7a}
    \end{subfigure}%
    ~
    \begin{subfigure}[t]{0.3\textwidth}
        \centering
        \includegraphics[width=6.2cm]{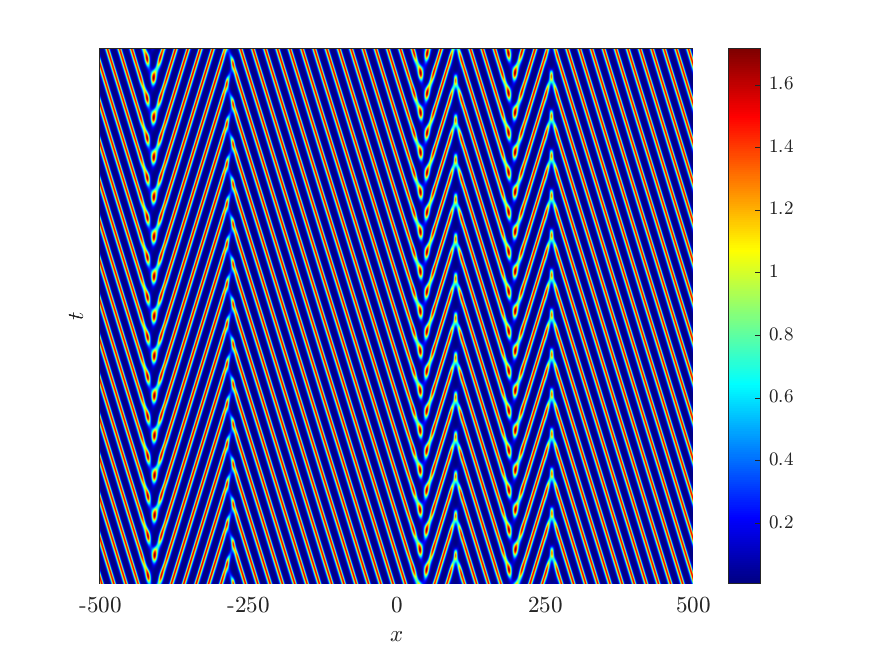}
        \caption{}\label{fig:7b}
    \end{subfigure}
    ~
    \begin{subfigure}[t]{0.3\textwidth}
        \centering
        \includegraphics[width=6.2cm]{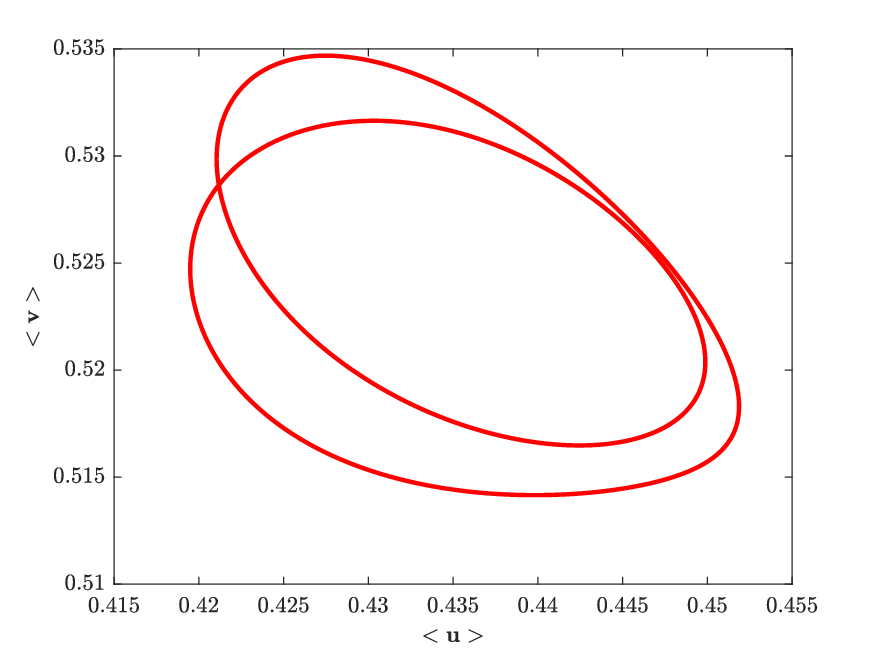}
        \caption{}\label{fig:7c}
    \end{subfigure}
    \caption{Formation of different types of patterns for $t\in [1000,1550]$ with $\sigma=10$ when $\omega$ is chosen from Hopf unstable domain and (a) $d_{2}=20$ (Turing- Hopf domain); and (b) $d_{2}=0.5$ (Hopf domain). (c) Spatial average of $u$ and $v$ for the solution plotted in (b) in Hopf domain for $t\in [30000, 35000 ]$.} \label{fig:7}
\end{figure}

Next, we consider $\sigma=10$ for which the curves $\Gamma_{\max}=0$ and $\Delta_{\min}=0$ intersect at a point. The curves $\Delta_{\min}=0$ and $\Gamma_{\max}=0$ together with the temporal Hopf curve divide the $\omega$-$d_{2}$ plane into six sub-regions [see Fig. \ref{fig:6c}]. In the pure spatial-Hopf domain, $\Gamma_{\max}>0$ and $\Delta_{\min}>0$ hold, and hence we find an oscillatory solution of the system (\ref{eq:loc1}). A wavetrain solution is plotted in Fig. \ref{fig:8a} for $d_2 = 21$, which oscillates in time. Since $\Gamma_{\max}<0$ and $\Delta_{\min}>0$ hold in the domain containing homogeneous steady solution, we get $u=u^{*}$ and $v = v^{*}$ of the nonlocal system. In the Turing spatial-Hopf domain, the conditions $\Gamma_{\max}>0$ and $\Delta_{\min}<0$ hold, and the system (\ref{eq:loc1}) gives oscillatory solutions or non-homogeneous steady patterns depending on the diffusion parameter values $d_{2}$. In this domain, the nonlocal model can produce oscillatory solutions in space and time, and such a solution is given in Fig. \ref{fig:8b}. The Turing domain corresponds to $\Gamma_{\max}<0$ and $\Delta_{\min}<0$, and an example of such a Turing pattern is plotted in Fig. \ref{fig:8c}. On the other hand, the curve $\Delta_{\min}=0$ divides the temporal Hopf unstable domain into two sub-domains: Turing- Hopf region where $\Delta_{\min}<0$ and Hopf region where $\Delta_{\min}>0$. We get oscillatory or non-homogeneous steady solutions in these regions. Figure \ref{fig:7} contains such solutions when $\omega=0.836(<\omega_{H})$. In particular, Fig. \ref{fig:7a} is portraited when $d_{2}(=20)$ lies in the Turing-Hopf region, whereas Fig. \ref{fig:7b} shows the solution for $d_{2}(=0.5)$ in Hopf domain.


\begin{figure}[htb!]
    \centering
    \begin{subfigure}[t]{0.3\textwidth}
        \centering
        \includegraphics[width=6.2cm]{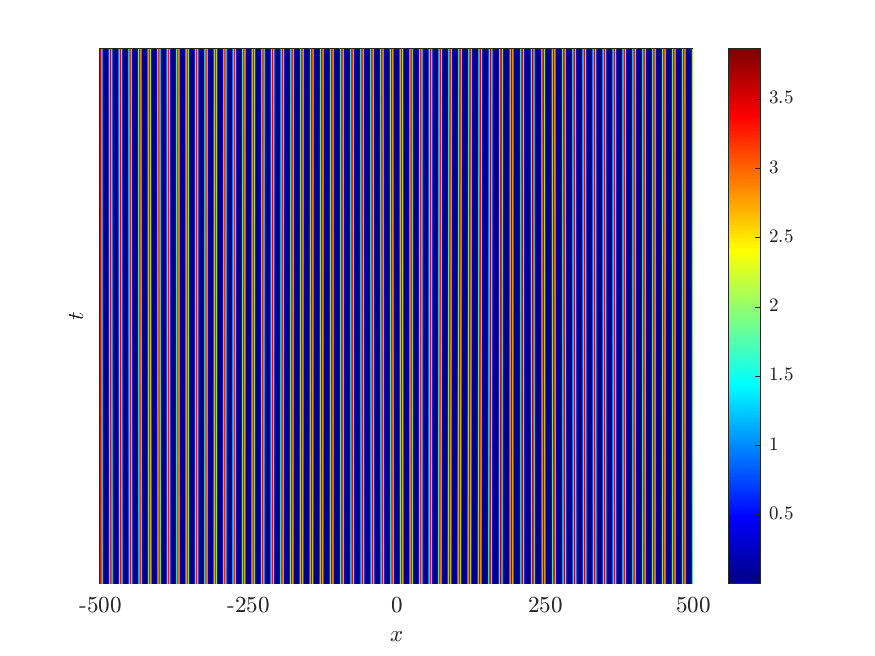}
        \caption{}\label{fig:9a}
    \end{subfigure}%
    ~
    \begin{subfigure}[t]{0.3\textwidth}
        \centering
        \includegraphics[width=6.2cm]{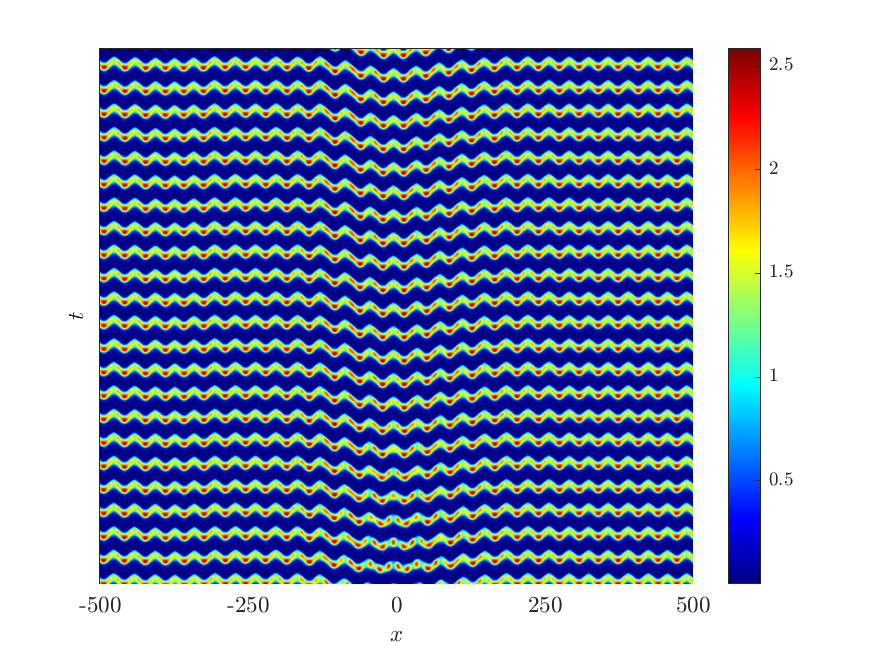}
        \caption{}\label{fig:9b}
    \end{subfigure}
    ~
    \begin{subfigure}[t]{0.3\textwidth}
        \centering
        \includegraphics[width=6.2cm]{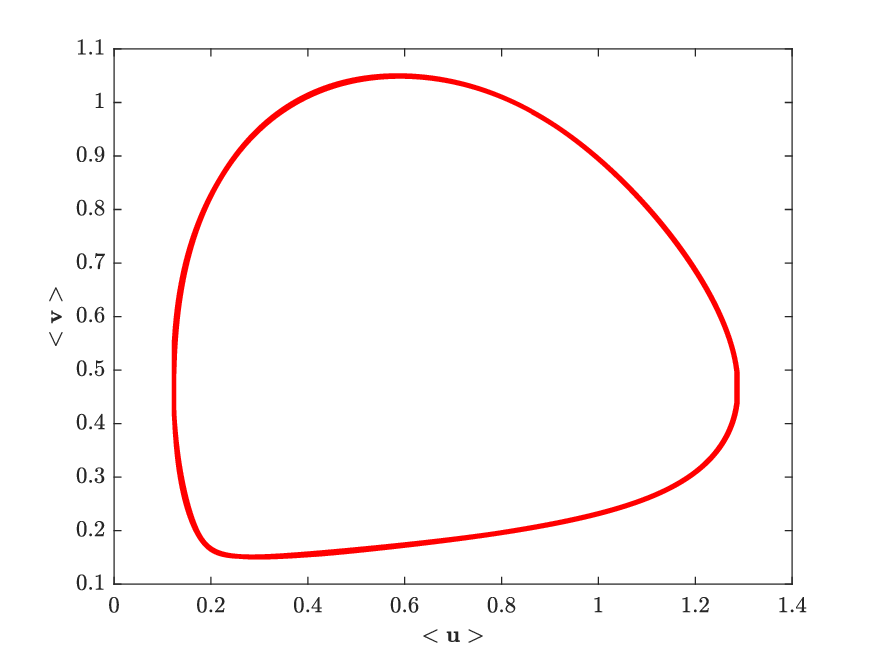}
        \caption{}\label{fig:9c}
    \end{subfigure}
    \caption{Different types of patterns for $\sigma=20$ when $\omega$ is chosen from Hopf unstable domain and (a) $d_{2}=20$ (Turing- Hopf domain); and (b) $d_{2}=1$ (Hopf domain). (c) Spatial average of $u$ and $v$ in Hopf domain for (b) for $t\in [30000, 33000]$.} \label{fig:9}
\end{figure}

\begin{figure}[htb!]
    \centering
    \begin{subfigure}[t]{0.3\textwidth}
        \centering
        \includegraphics[width=6.2cm]{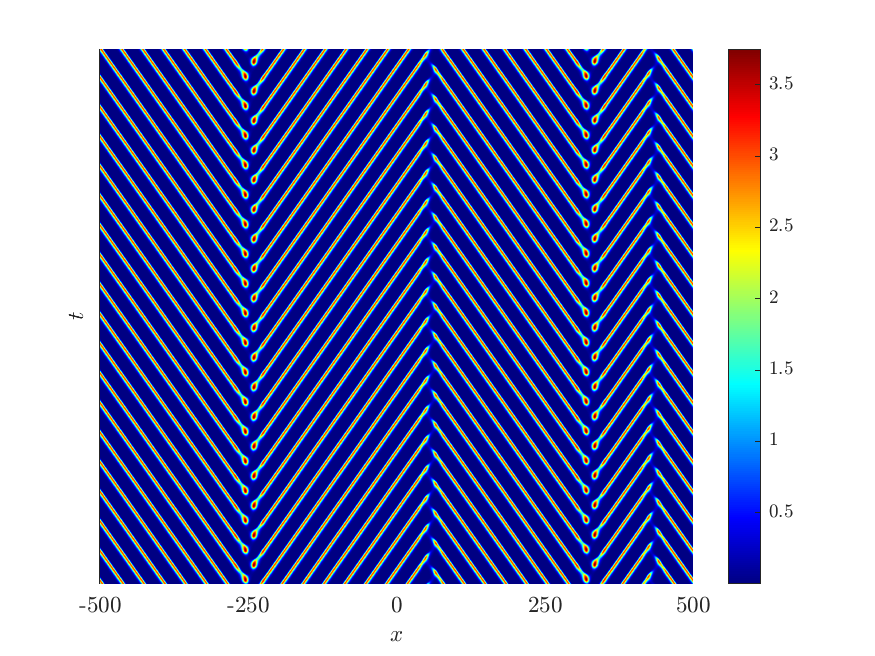}
        \caption{}\label{fig:10a}
    \end{subfigure}%
    ~
    \begin{subfigure}[t]{0.3\textwidth}
        \centering
        \includegraphics[width=6.2cm]{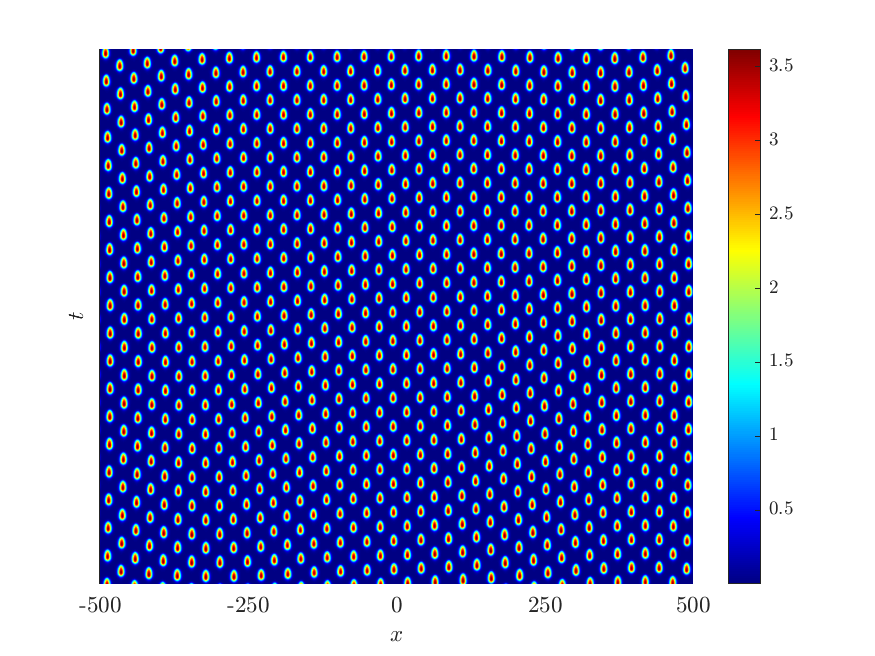}
        \caption{}\label{fig:10b}
    \end{subfigure}
    ~
    \begin{subfigure}[t]{0.3\textwidth}
        \centering
        \includegraphics[width=6.2cm]{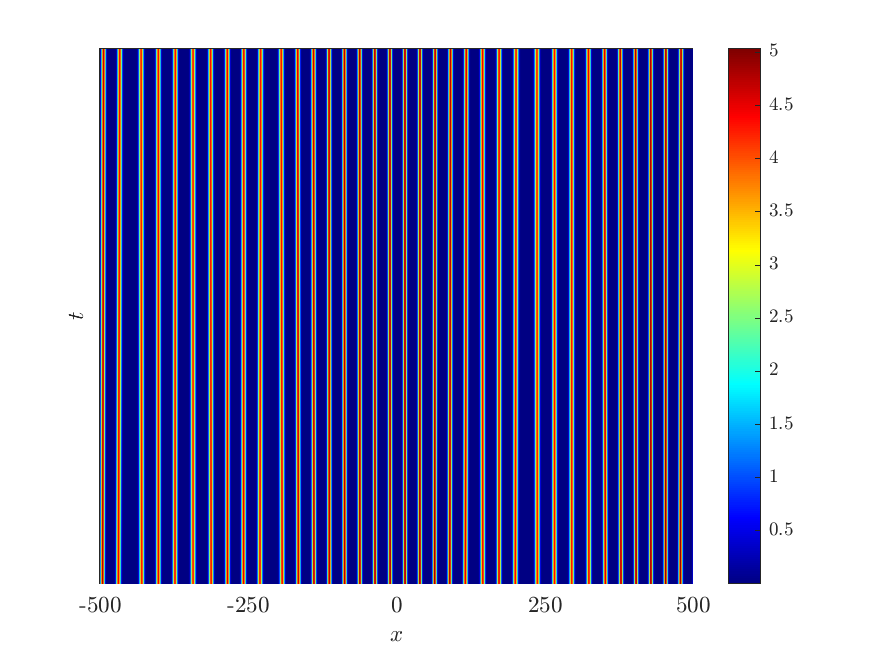}
        \caption{}\label{fig:10c}
    \end{subfigure}
    \caption{Color plot of $u$ for $t\in[1000,1550]$ with $\sigma=20$ when $\omega$ is chosen from Hopf stable domain and (a) $d_{2}(=1)<d_{2c}^{N}$ (Pure spatial-Hopf domain), (b) $d_{2c}^{N}<d_{2}(=5)<d_{2c}^{H}$ (Turing spatial-Hopf domain), (c) $d_{2}(=25)>d_{2c}^{H}$ (Turing domain).} \label{fig:10}
\end{figure}

Lastly, we consider $\sigma=20$ for which the curve $\Delta_{\min}=0$ lies below the curve $\Gamma_{\max}=0$. The curves $\Gamma_{\max}=0, \ \Delta_{\min}=0$ and the Hopf curve divide the considered region of the $\omega$-$d_{2}$ plane into five sub-regions [see Fig. \ref{fig:6d}]. As the Turing curve divides the Hopf unstable domain into Turing- Hopf and Hopf region, we get oscillatory or non-homogeneous steady solutions there. Figure \ref{fig:9a} is shown when $d_{2}(=20)$ is chosen from the Turing-Hopf region, whereas Fig. \ref{fig:9b} shows the solution for $d_{2}(=1)$ in Hopf domain. The conditions $\Gamma_{\max}>0$ and $\Delta_{\min}>0$ hold in the pure spatial-Hopf domain, which gives occurrence of oscillatory or stationary patterns [see Fig. \ref{fig:10a}]. We have $\Gamma_{\max}>0$ and $\Delta_{\min}<0$ in the Turing spatial-Hopf domain, and hence, we observe the non-homogeneous steady patterns [see Fig. \ref{fig:10b}]. It is observed that the oscillatory solutions occur when $d_{2}$ lies in the Turing spatial-Hopf domain, but stationary patterns dominate the oscillatory solutions when the parameter $d_{2}$ lies near the spatial-Hopf curve. Turing instability conditions $\Gamma_{\max}<0$ and $\Delta_{\min}<0$ are satisfied in the Turing domain, which leads to a stationary Turing pattern [see Fig. \ref{fig:10c}]. It is noted that there does not exist any domain containing a homogeneous solution in $\omega$-$d_{2}$ plane for $\sigma=20$, but oscillatory solution patterns in the pure spatial-Hopf region are shown in Fig. \ref{fig:10a}. 

\begin{figure}[htb!]
    \centering
    \begin{subfigure}[t]{0.5\textwidth}
        \centering
        \includegraphics[width=7.5cm,height=5.2cm]{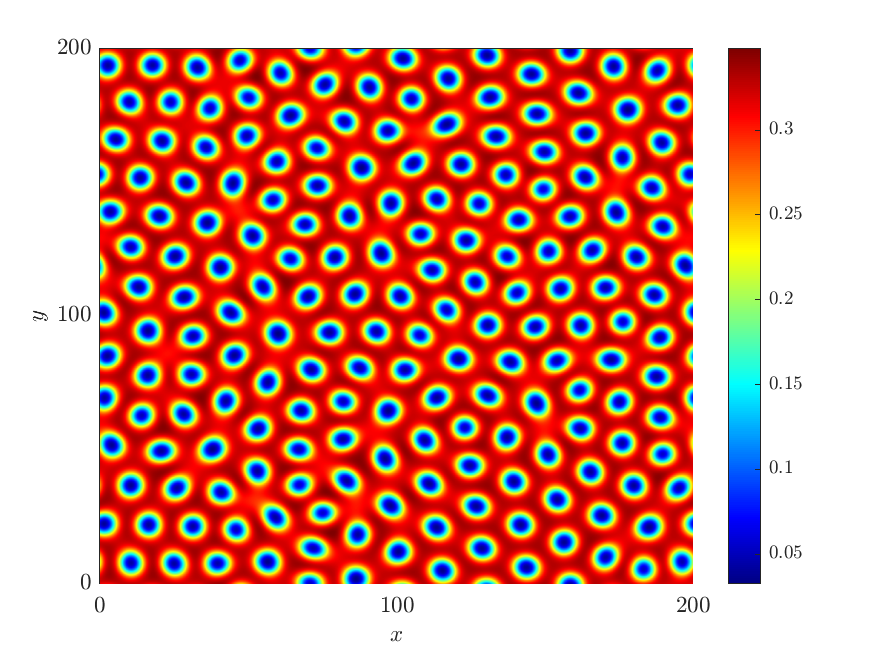}
        \caption{$(\omega,\ d_{2})=(0.85,60)$}\label{fig:12a}
    \end{subfigure}%
    ~
    \begin{subfigure}[t]{0.5\textwidth}
        \centering
        \includegraphics[width=7.5cm,height=5.2cm]{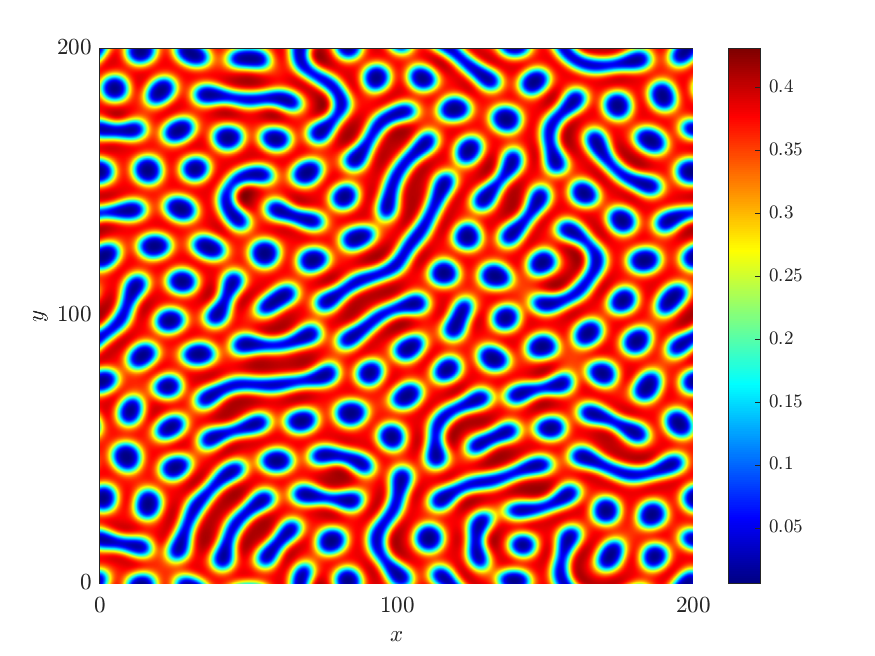}
        \caption{$(\omega,\ d_{2})=(0.8,80)$}\label{fig:12b}
    \end{subfigure}

    \begin{subfigure}[t]{0.5\textwidth}
        \centering
        \includegraphics[width=7.5cm,height=5.2cm]{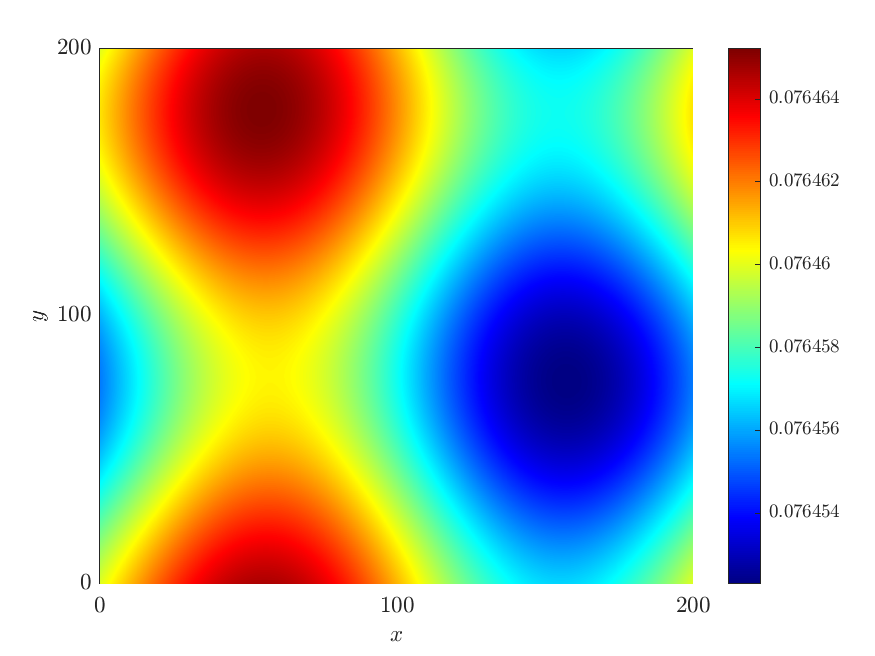}
        \caption{$(\omega,\ d_{2})=(0.8,10)$}\label{fig:12c}
    \end{subfigure}%
    ~
    \begin{subfigure}[t]{0.5\textwidth}
        \centering
        \includegraphics[width=7.5cm,height=5.2cm]{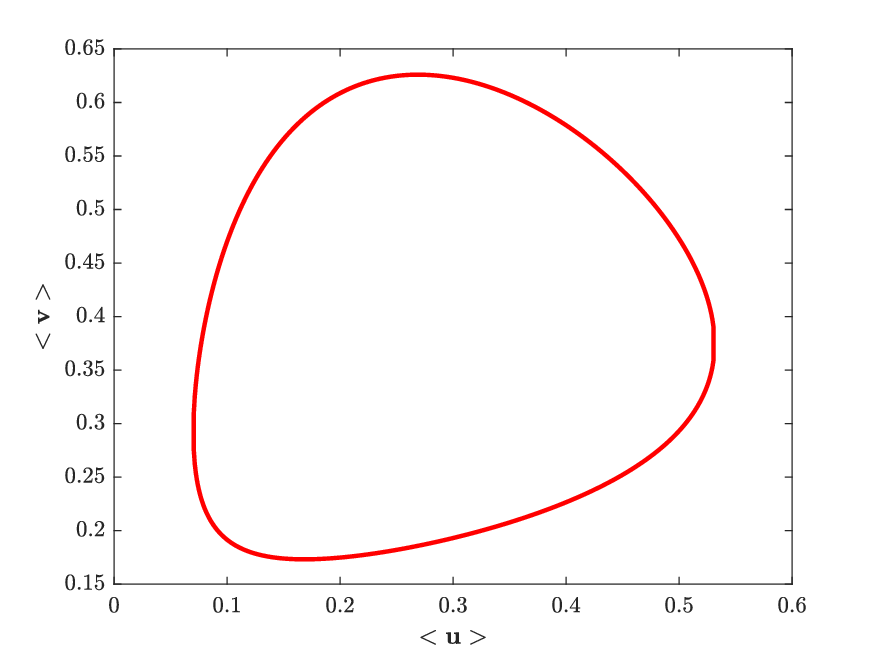}
        \caption{Spatial average of $u$ and $v$  for $t\in [15000, 18000 ]$.}\label{fig:12d}
    \end{subfigure}
    \caption{Solutions for the prey species in two dimensions for different values of $(\omega, d_{2})$. The spatial domain is chosen as $[0,200]\times[0,200]$ with $dx=dy=0.5$ and $dt=0.0005$.} \label{fig:12}
\end{figure}

The contour plot of the prey species for the spatio-temporal model is plotted in Fig. \ref{fig:4} when $(\omega, d_{2})$ is chosen from Turing, Turing-Hopf and Hopf domain. It is observed that there are not that many qualitative changes in the solutions. Here, we discuss two examples of patterns in two spatial dimensions whose patterns in one spatial dimension look the same. Figure \ref{fig:12} depicts such patterns for the parameter values $(\omega, d_2) = (0.85,60)$ and $(\omega, d_2) = (0.8,80)$, and the corresponding patterns in one dimension are shown in Fig. \ref{fig:4}. Therefore, the two spatial dimensions give more diversity in the stationary patterns. So, it would be interesting to consider the nonlocal interaction in a two-dimensional spatial domain in the future and see the pattern formations when the parameters $\omega$ and $d_{2}$ reside in distinct domains for different ranges of nonlocal interactions. 

Overall, the numerical simulations yield numerous critical insights into the proposed predator-prey system with inducible prey defence. We have demonstrated that predator numbers continue to climb while defence is minimal, but when defence rises, the prey population grows significantly, resulting in a fall in predator numbers. The temporal study reveals that inducible defence has a stabilizing influence on population counts. Furthermore, in the spatio-temporal model, defence lowers the size of the Turing domain, preventing spatial pattern formation. However, when a nonlocal term is included, the Turing domain gets bigger, allowing for complex spatial patterns. Depending on the value of the predator's diffusion coefficient in the presence of the nonlocal term, different outcomes, such as stationary, non-stationary, and oscillatory patterns, can emerge in the model system.

\section{Conclusions} \label{sec:6}

A kind of phenotypic plasticity known as inducible defences has the ability to alter direct interactions between different members of an ecological community, leading to trait-mediated indirect effects \cite{wootton2002indirect, bolker2003connecting, werner2003review}. Direct contact between prey and predator is inhibited when a prey exhibits a protective behavioural, morphological, or physiological feature in response to predator density. Moreover, when the defence has costs, the prey's development rate is slowed down (metabolic costs), or the prey-resource interaction is inhibited (feeding costs). Inducible defence can thus have trait-mediated impacts on the predator, its prey, and the prey's resource through a change of interactions \cite{wootton1993indirect}. These defences are preferred over constitutive ones if the protective characteristic costs the prey, according to evolutionary ecological theory \cite{harvell1990ecology, harvell1999inducible}. There are already some predator-prey interactions are studied with the implications of inducible defence at the population and community levels \cite{ives1987antipredator, abrams1996invulnerable, ramos2000relating, ramos2002population, ramos2003population, kopp2006dynamic}. However, our goal in this study is to comprehend how this inducible defence impacts the dynamical scenario of predator-prey interaction in the presence of nonlocal interaction. \par

In this work, we have come up with a system with the involvement of prey's inducible defence strategy. The incorporation of the defence has significantly increased the count of prey species in the system. Not only that, this defence is proven to be a stabilizing component for the system as increasing the level of defence helps to coexist the species in the environment as a stable steady-state. Instead of analyzing the temporal model only, we intend to focus on the involvement of species diffusion in the model. The species are assumed to move in a one- and two-dimensional bounded domain. The result shows that the Turing domain starts to shrink with the increase of inducible defence level leading to a reduction in species colonization. The numerical simulation shows that the system exhibits cold spots when $\omega$ is chosen from the Hopf stable domain and $d_{2}>d_{2c}$ (Turing domain), but mixed pattern when chosen from the Hopf unstable domain (Turing-Hopf domain). \par

The work is further extended into a nonlocal model by incorporating a nonlocal intra-specific competition in prey species. It is shown that the size of the Turing domain changes with the implementation of nonlocal interaction. In Fig. \ref{fig:6}, the spatial-Hopf curve is also portrayed that appears in the Hopf stable region. From Fig. \ref{fig:3} it is obtained that the Hopf curve and Turing curve divide the $\omega-d_{2}$ plane into four regions, while on the other hand, incorporation of nonlocal term causes the occurrence of the spatial-Hopf curve because of which the $\omega-d_{2}$ plane is divided into six different regions where homogeneous and non-homogeneous solutions can be observed and this is reflected in Fig. \ref{fig:6}. \par

The Hopf and Turing curves split the $\omega$-$d_{2}$ plane area into four different zones for low nonlocal interaction ranges. As the nonlocal interaction range expands, the Turing curve changes downward, resulting in larger Turing and Turing-Hopf domains and smaller Hopf and stable areas. Stationary non-homogeneous patterns thus occur in more extensive regions as these are predominantly seen in the Turing and Turing Hopf domains. In contrast, the stationary homogeneous states and oscillatory solutions contract in size. As nonlocal interaction ranges rise, a spatial Hopf curve forms in the stable zone. As a result, the stable zone becomes smaller. The solution in the spatial Hopf zone shows non-homogeneous oscillations. When the spatial Hopf curve is below the Turing curve, areas with stationary non-homogeneous solutions rise, whereas stable and homogeneous oscillatory solutions diminish as the range of nonlocal interaction increases. As nonlocal interaction ranges expand, the spatial Hopf curve rises above the Turing curve, and the stable area vanishes. This decreases the Turing domain while increasing the spatial Hopf domain. The areas of non-homogeneous oscillatory solution grow. Our findings indicate that nonlocal interactions have a considerable impact on the spatiotemporal properties of a prey-predator model. This might have ramifications for ecological prey-predator models. \par

Inducible defences have the potential to stabilize the food chain through two different processes. First, protective features impair the focused predator-prey contact by reducing the functional response. Furthermore, the expenses either decrease the prey growth rate (metabolic costs) or erode the prey-resource relationship (feeding costs). According to McCann et al., one of the stabilizing processes in food webs is the weakening of trophic connections \cite{mccann1998weak}. Conversely, predator abundance generates negative feedback on itself, signalling a decline in contact intensity \cite{kopp2006dynamic}. It is widely accepted that this self-regulating mechanism of negative feedback is a powerful stabilizing factor in population systems \cite{puccia1985qualitative, berryman2020principles, dambacher2007understanding}. Consequently, trait-mediated indirect effects in general and ID in particular may undermine interactions, leading to damping oscillations and boosting community stability. \par

Despite having rich dynamics, the suggested system can yet be improved in the future. One of the restrictions under which the system is designed is the assumption that the inducible defence of prey species influences their development. However, as a countermeasure, the predator species may use alternative hunting techniques in their natural habitat. Thus, considering the role of several functional responses in both the predator's and prey's development will bring the situation closer to reality. Furthermore, the carryover effect can occur in any predator-prey relationship in ecological systems where a species' background and experiences are utilised to explain its current conduct. One type of phenotypic plasticity that alters the direct interactions between different members of an ecological community and produces trait-mediated indirect consequences is this inducible defence of prey. Therefore, it could impact several generations rather than just one. It suggests that the prey species in the model can incorporate the carryover effect. Additionally, ambient stochasticity can be introduced into the system via white Gaussian noise, giving even more realistic assumptions in the current framework. \par

Currently, there is interest in considering the indirect impacts of changes in interactions to comprehend community behaviour \cite{werner2003review, dambacher2007understanding}. Nonetheless, examining how these intricate relationships contribute to the spread, mitigation, or intensification of the detrimental effects of environmental stressors on the outcomes of populations enmeshed in actual and simulated communities may yield a wealth of new information \cite{fleeger2003indirect,rohr2006community}. Theoretical and experimental studies addressing the effects of interaction changes on the structure and operation of ecological networks will undoubtedly be helpful for further research.

\subsection{Acknowledgements}
The authors thank the NSERC and the CRC Program for their support. RM also acknowledges the support of the BERC 2022–2025 program and the Spanish Ministry of Science, Innovation and Universities through the Agencia Estatal de Investigacion (AEI) BCAM Severo Ochoa excellence accreditation SEV-2017–0718. This research was enabled in part by support provided by SHARCNET (\url{www.sharcnet.ca}) and Digital Research Alliance of Canada (\url{www.alliancecan.ca}).

\subsection{Data Availability Statement}
The data used to support the findings of the study are available within the article.

\subsection{Conflict of Interest}
This work does not have any conflict of interest.

\bibliographystyle{elsarticle-num} 
\bibliography{P_References}

\end{document}